%
%
\let\includefigures=\iftrue
%


\input harvmac
\newcount\yearltd\yearltd=\year\advance\yearltd by 0



\input epsf

\newcount\figno
\figno=0
\def\fig#1#2#3{
\par\begingroup\parindent=0pt\leftskip=1cm\rightskip=1cm\parindent=0pt
\baselineskip=11pt
\global\advance\figno by 1
\midinsert
\epsfxsize=#3
\centerline{\epsfbox{#2}}
\vskip 12pt
{\bf Figure \the\figno:} #1\par
\endinsert\endgroup\par
}
\def\figlabel#1{\xdef#1{\the\figno}}


\noblackbox
\def\IZ{\relax\ifmmode\mathchoice
{\hbox{\cmss Z\kern-.4em Z}}{\hbox{\cmss Z\kern-.4em Z}}
{\lower.9pt\hbox{\cmsss Z\kern-.4em Z}} {\lower1.2pt\hbox{\cmsss
Z\kern-.4em Z}}\else{\cmss Z\kern-.4em Z}\fi}

\font\cmss=cmss10 \font\cmsss=cmss10 at 7pt
\def\IR{\relax{\rm I\kern-.18em R}}

\def\frac#1#2{{#1 \over #2}}
\def\cn{{\cal N}}
\def\th{{\tilde h}}
\def\tth{{\tilde \theta}}
\def\tJ{{\tilde J}}
\def\bdel{{\bar \del}}

\def\tq{{\tilde q}}

\noblackbox

\lref\AharonyPA{ O.~Aharony, M.~Berkooz and E.~Silverstein,
``Multiple-trace operators and non-local string theories,''
JHEP {\bf 0108}, 006 (2001) [arXiv:hep-th/0105309].
}

\lref\KutasovXU{
D.~Kutasov and N.~Seiberg,
``More comments on string theory on $AdS_3$,''
JHEP {\bf 9904}, 008 (1999)
[arXiv:hep-th/9903219].
}

\lref\deBoerPP{
J.~de Boer, H.~Ooguri, H.~Robins and J.~Tannenhauser,
``String theory on $AdS_3$,''
JHEP {\bf 9812}, 026 (1998)
[arXiv:hep-th/9812046].
}

\lref\KachruYS{
S.~Kachru and E.~Silverstein,
``4d conformal theories and strings on orbifolds,''
Phys.\ Rev.\ Lett.\  {\bf 80}, 4855 (1998)
[arXiv:hep-th/9802183].
}

\lref\GiveonNS{
A.~Giveon, D.~Kutasov and N.~Seiberg,
``Comments on string theory on $AdS_3$,''
Adv.\ Theor.\ Math.\ Phys.\  {\bf 2}, 733 (1998)
[arXiv:hep-th/9806194].
}

\lref\usnext{O. Aharony, M. Berkooz, E. Silverstein, ... work in
progress.}

\lref\MaldacenaRE{
J.~Maldacena,
``The large $N$ limit of superconformal field theories and supergravity,''
Adv.\ Theor.\ Math.\ Phys.\  {\bf 2}, 231 (1998)
[Int.\ J.\ Theor.\ Phys.\  {\bf 38}, 1113 (1998)]
[arXiv:hep-th/9711200].
}

\lref\DijkgraafJT{
R.~Dijkgraaf, E.~Verlinde and H.~Verlinde,
``On Moduli Spaces Of Conformal Field Theories With c $\geq$ 1,''
{\it in *Copenhagen 1987, proceedings, perspectives in string theory*,
117-137. }
}

\lref\BalasubramanianNH{
V.~Balasubramanian, M.~Berkooz, A.~Naqvi and M.~J.~Strassler,
``Giant gravitons in conformal field theory,''
[arXiv:hep-th/0107119].
}

\lref\LarsenUK{
F.~Larsen and E.~J.~Martinec,
``$U(1)$ charges and moduli in the D1-D5 system,''
JHEP {\bf 9906}, 019 (1999)
[arXiv:hep-th/9905064].
}

\lref\GiveonUQ{
A.~Giveon, D.~Kutasov and A.~Schwimmer,
``Comments on D-branes in $AdS_3$,''
Nucl.\ Phys.\ B {\bf 615}, 133 (2001)
[arXiv:hep-th/0106005].
}

\lref\WittenQJ{
E.~Witten,
``Anti-de Sitter space and holography,''
Adv.\ Theor.\ Math.\ Phys.\  {\bf 2}, 253 (1998)
[arXiv:hep-th/9802150].
}

\lref\GubserBC{
S.~S.~Gubser, I.~R.~Klebanov and A.~M.~Polyakov,
``Gauge theory correlators from non-critical string theory,''
Phys.\ Lett.\ B {\bf 428}, 105 (1998)
[arXiv:hep-th/9802109].
}

\lref\AharonyTI{
O.~Aharony, S.~S.~Gubser, J.~Maldacena, H.~Ooguri and Y.~Oz,
``Large $N$ field theories, string theory and gravity,''
Phys.\ Rept.\  {\bf 323}, 183 (2000)
[arXiv:hep-th/9905111].
}

\lref\BalasubramanianRT{
V.~Balasubramanian, J.~de Boer, E.~Keski-Vakkuri and S.~F.~Ross,
``Supersymmetric conical defects: Towards a string theoretic description
of black hole formation,''
Phys.\ Rev.\ D {\bf 64}, 064011 (2001)
[arXiv:hep-th/0011217].
}

\lref\WittenHF{
E.~Witten,
``Quantum Field Theory And The Jones Polynomial,''
Commun.\ Math.\ Phys.\  {\bf 121}, 351 (1989).
}

\lref\ElitzurNR{
S.~Elitzur, G.~W.~Moore, A.~Schwimmer and N.~Seiberg,
``Remarks On The Canonical Quantization Of The Chern-Simons-Witten Theory,''
Nucl.\ Phys.\ B {\bf 326}, 108 (1989).
}

\lref\MaldacenaSS{
J.~Maldacena, G.~W.~Moore and N.~Seiberg,
``D-brane charges in five-brane backgrounds,''
JHEP {\bf 0110}, 005 (2001)
[arXiv:hep-th/0108152].
}

\lref\GiveonUQ{
A.~Giveon, D.~Kutasov and A.~Schwimmer,
``Comments on D-branes in $AdS_3$,''
Nucl.\ Phys.\ B {\bf 615}, 133 (2001)
[arXiv:hep-th/0106005].
}

\lref\BachasFR{
C.~Bachas and M.~Petropoulos,
``Anti-de-Sitter D-branes,''
JHEP {\bf 0102}, 025 (2001)
[arXiv:hep-th/0012234].
}

\lref\RajaramanEW{
A.~Rajaraman,
``New $AdS_3$ branes and boundary states,''
arXiv:hep-th/0109200.
}

\lref\ParnachevGW{
A.~Parnachev and D.~A.~Sahakyan,
``Some remarks on D-branes in $AdS_3$,''
JHEP {\bf 0110}, 022 (2001)
[arXiv:hep-th/0109150].
}

\lref\RajaramanCR{
A.~Rajaraman and M.~Rozali,
``Boundary states for D-branes in $AdS_3$,''
arXiv:hep-th/0108001.
}

\lref\HikidaYI{
Y.~Hikida and Y.~Sugawara,
``Boundary states of D-branes in $AdS_3$ based on discrete series,''
arXiv:hep-th/0107189.
}

\lref\LeeXE{
P.~Lee, H.~Ooguri, J.~W.~Park and J.~Tannenhauser,
``Open strings on $AdS_2$ branes,''
Nucl.\ Phys.\ B {\bf 610}, 3 (2001)
[arXiv:hep-th/0106129].
}

\lref\PetropoulosQU{
P.~M.~Petropoulos and S.~Ribault,
``Some remarks on anti-de Sitter D-branes,''
JHEP {\bf 0107}, 036 (2001)
[arXiv:hep-th/0105252].
}

\lref\StanciuNX{
S.~Stanciu,
``D-branes in an $AdS_3$ background,''
JHEP {\bf 9909}, 028 (1999)
[arXiv:hep-th/9901122].
}

\lref\FigueroaOFarrillEI{
J.~M.~Figueroa-O'Farrill and S.~Stanciu,
``D-branes in $AdS_3\times S^3\times S^3\times S^1$,''
JHEP {\bf 0004}, 005 (2000)
[arXiv:hep-th/0001199].
}

\lref\RyangPX{
S.~J.~Ryang,
``Nonstatic $AdS_2$ branes and the isometry group of $AdS_3$ spacetime,''
arXiv:hep-th/0110008.
}

\lref\MaldacenaKY{
J.~Maldacena, G.~W.~Moore and N.~Seiberg,
``Geometrical interpretation of D-branes in gauged WZW models,''
JHEP {\bf 0107}, 046 (2001)
[arXiv:hep-th/0105038].
}

\lref\GiddingsCX{
S.~B.~Giddings and A.~Strominger,
``Loss Of Incoherence And Determination
Of Coupling Constants In Quantum Gravity,''
Nucl.\ Phys.\ B {\bf 307}, 854 (1988).
}

\lref\GiddingsWV{
S.~B.~Giddings and A.~Strominger,
``Baby Universes, Third Quantization And The Cosmological Constant,''
Nucl.\ Phys.\ B {\bf 321}, 481 (1989).
}

\lref\ColemanTJ{
S.~R.~Coleman,
``Why There Is Nothing Rather Than Something:
A Theory Of The Cosmological Constant,''
Nucl.\ Phys.\ B {\bf 310}, 643 (1988).
}

\lref\HawkingMZ{
S.~W.~Hawking,
``Quantum Coherence Down The Wormhole,''
Phys.\ Lett.\ B {\bf 195}, 337 (1987).
}

\lref\MaldacenaKM{
J.~Maldacena and H.~Ooguri,
``Strings in $AdS_3$ and the $SL(2,\IR)$ WZW
model. III: Correlation  functions,''
arXiv:hep-th/0111180.
}


\def\myTitle#1#2{\nopagenumbers\abstractfont\hsize=\hstitle\rightline{#1}%
\vskip 0.5in\centerline{\titlefont #2}\abstractfont\vskip .4in\pageno=0}

\myTitle{\vbox{\baselineskip12pt\hbox{hep-th/0112178}
\hbox{SLAC-PUB-9099}
\hbox{WIS/28/01-DEC-DPP}
}} {\vbox{
\centerline{Non-Local String Theories on $AdS_3\times S^3$}
\medskip
\centerline{and Stable Non-Supersymmetric Backgrounds}}}
\centerline{Ofer Aharony$^{a,}$\foot{E-mail :
{\tt Ofer.Aharony@weizmann.ac.il.}
Incumbent of the Joseph and Celia Reskin
career development chair.}, Micha Berkooz$^{a,}$\foot{E-mail : {\tt
Micha.Berkooz@weizmann.ac.il.}} and Eva
Silverstein$^{b,}$\foot{E-mail : {\tt evas@slac.stanford.edu.}}}
\medskip
\centerline{$^{a}$Department of Particle Physics, The Weizmann
Institute of Science, Rehovot 76100, Israel}
\medskip
\centerline{$^{b}$Department of Physics and SLAC, Stanford
University, Stanford, CA 94305/94309, USA}

\bigskip
\noindent

We exhibit a simple class of exactly marginal ``double-trace''
deformations of two dimensional CFTs which have $AdS_3$ duals, in
which the deformation is given by a product of left and right-moving
$U(1)$ currents. In this special case the deformation on $AdS_3$ is generated
by a local boundary term in three dimensions, which changes the
physics also in the bulk via bulk-boundary propagators.
However, the deformation
is non-local in six dimensions and on the string worldsheet,
like generic non-local string theories (NLSTs).
Due to the simplicity of the deformation we can explicitly
make computations in the non-local string theory and compare them to
CFT computations, and we obtain precise agreement. We discuss the
effect of the deformation on closed strings and on D-branes. The
examples we analyze include a supersymmetry-breaking but exactly
marginal ``double-trace'' deformation, which is dual to a string
theory in which no destabilizing tadpoles are generated for moduli
nonperturbatively in all couplings, despite the absence of
supersymmetry. We explain how this cancellation works on the gravity
side in string perturbation theory, and also non-perturbatively at
leading order in the deformation parameter. We also discuss possible
flat space limits of our construction.

\Date{December 2001}

\baselineskip=16pt

\newsec{Introduction}

One interesting direction of research in string/M theory concerns novel phases
of the theory. Examples include non-commutative Yang-Mills theory
and non-geometrical phases of string compactifications. Although
such phases may appear to be exotic, in some cases they are generic, in
the sense that returning to more conventional backgrounds requires
tuning a superselection parameter to a special value. These novel
backgrounds are very much worth studying, both because of their
intrinsic interest and because of the hope that their
unconventional physics may play a role in solving open problems
that remain in formulating and applying the theory (such as the
cosmological constant problem).

In \AharonyPA\ we found strong evidence for a new type of
perturbative string theory, non-local string theory (NLST),
arising on the gravity side of AdS/CFT \refs{\MaldacenaRE,\WittenQJ,
\GubserBC,\AharonyTI} dual pairs whose
field theory side is deformed by a ``multi-trace''
operator\foot{We will use the names ``single-trace'' and
``multi-trace'' operators for any CFT which has a weakly-curved
AdS dual, though the operators can only be represented in terms of
traces in the case of four dimensional gauge theories. By a
``single-trace'' operator we will mean an operator which is dual
to a single particle in string theory (for example, a KK mode of
the graviton), while ``multi-trace'' operators will appear in the
OPE of such operators. The distinction between these classes of
operators is not always clear (see, e.g., \BalasubramanianNH), but
it can be made in an obvious way for operators of low dimension
when the background is weakly curved (such ``single-trace''
operators correspond simply to supergravity fields) and this is
all that we will use here.}. In such theories, the ``exotic'' phase is
generic, since it is obvious on the field theory side of the duality
that one has to tune parameters in order to get back to the
conventional theory, so the conventional string theory occupies a
set of measure zero in the space of theories.  These theories are
gravitational, and have many intriguing features outlined in
\AharonyPA. In a perturbative string description, the perturbative
expansion in the deformation is reproduced by shifting the
worldsheet action by a bilocal term of the general form
\eqn\genwsdef{ \delta S_{ws}=\sum_{I, J} {\tilde h}_{IJ}\int
d^2z_1 V^{(I)}[y(z_1)] \int d^2z_2 V^{(J)}[y(z_2)], } where
$V^{(I)}$ are some vertex operators in the string theory each
including a factor of the string coupling $g_s$ (in the
examples of \AharonyPA\ the index $I$ was continuous), and $y(z)$
are the embedding coordinates of the string worldsheet (or any
other fields on the worldsheet). In \AharonyPA\ examples of
double-trace deformations which were relevant or marginal in the dual
CFT were exhibited. It
was shown that these deformations could not be accounted for by
local 10-dimensional supergravity, and that, in perturbation
theory in the strength $\tilde h$ of the deformation, the changes
in CFT correlators are formally reproduced by the shift \genwsdef\
in the worldsheet action.  This leads to a new type of
diagrammatic expansion encoding the perturbation theory in both
$\tilde h$ and $g_s$ which has many interesting novel
features.  In particular, at a given order $n$ in the $g_s$ expansion,
one has contributing diagrams which do not have the modular properties of
genus $n$ Riemann surfaces.

In these theories, some sectors are affected by the deformation
at leading order
in $g_s$ (classically on the gravity side), while other
sectors are not.  For instance, exclusive graviton scattering
along the $AdS$ directions remains the same at tree level
on the gravity side \AharonyPA.  This parametric hierarchy
between an approximately
local sector and a completely non-local sector
for small string coupling on the gravity
side may potentially render these theories more viable as
physical models than they would be otherwise.

The examples of \AharonyPA\ involved string theory in RR backgrounds,
so it was difficult to make
the formal expression \genwsdef\
more explicit, due to the current limitations on our understanding
of RR backgrounds in string theory.  It is important
to study more explicitly the conformal perturbation expansion around
the undeformed background, in order to understand how divergences
arising in conformal perturbation theory are regularized from
the point of view of both sides of the duality, and
in order to make progress on the larger questions regarding the
consistency, degree of non-locality, and applications of the new theories.

In this paper, we present a rather explicit example of an interesting
``double-trace'' deformation in the Neveu-Schwarz version of
$AdS_3/CFT_2$ arising from the low energy/near horizon
limit of a system of $Q_1$ fundamental strings and $Q_5$ NS 5-branes
\MaldacenaRE.
In the dual CFT this deformation is of the form $\delta
S_{CFT} \simeq {{\tilde h\over{Q_1Q_5}}} \int d^2x J(x)
\tJ(\bar x)$ where $J$ and $\tilde J$ are left and right moving
global symmetry currents in the dual CFT.
By using the explicit string theory description of undeformed
$AdS_3/CFT_2$ that has been developed in recent years
(see for example the comprehensive analysis in \MaldacenaKM\
and references therein) -- in particular the formalism of
\refs{\GiveonNS,\KutasovXU} for vertex operators and correlation
functions and the semiclassical analysis of \deBoerPP\ -- we are able
to analyze explicitly many aspects of this deformation.  In
particular, we check explicitly the absorption of divergences in
conformal perturbation theory.

This deformation has an interesting physical property.  It
is exactly marginal but at the same time, if $J$ and $\tJ$
are $U(1)$ currents in the R-symmetry group, it
breaks supersymmetry.  Applying the basic relation between conformal
invariance and AdS isometries \MaldacenaRE\ to nonsupersymmetric
systems leads to an interesting element in the duality dictionary \KachruYS.
Namely, when there is a non-supersymmetric
hypersurface of RG fixed points,
a destabilizing potential
for moduli is not generated along this hypersurface despite the
absence of supersymmetry.

Our model provides for the first time an example realizing this
possibility where the fixed surface exists for finite values of
the string coupling. The price of this (which may end up being a
positive feature) is that the fixed surface includes a
``double-trace'' deformation which controls the strength of supersymmetry
breaking. Perturbatively in the string coupling $g_s$, and also
non-perturbatively in $g_s$ at first
order in $\tilde h$, we find a simple cancellation mechanism that
reproduces the cancellation of the moduli potential directly on
the gravity side. For higher orders in $\tilde h$ we do not yet
understand directly the way the cancellation occurs beyond
string perturbation theory on the gravity
side; this is a very intriguing prediction of the duality. The
supersymmetry breaking in this model is ``hard'', in that the
supersymmetry-breaking splittings of the masses (which are related to
the splittings between the dimensions of
corresponding operators in the dual CFT) grow with the masses.
Unfortunately, the supersymmetry breaking effects
are small -- they disappear when
we take the flat space limit, so that this does not yet provide a
basis for a realistic theory of supersymmetry breaking. However,
the cancellation of tadpoles for moduli is nontrivial in our model
for finite $AdS$ radius, since the (vanishing) moduli tadpoles are
hierarchically smaller than the scale of supersymmetry breaking.

Given this prediction for stability after supersymmetry
breaking, and more generally in the interest of
clarifying the physics of NLST's, it is important to study the effects of
the deformation on bulk physics on the gravity side of
the correspondence.

The deformation has interesting effects on both the perturbative and
non-perturbative sectors of the theory.
The dimensions of operators corresponding to charged
particles propagating in AdS are changed by
the deformation.  As far as the perturbative
sector is concerned, because the ``double-trace'' deformation in this
specific case involves vertex operators which are total derivatives on
the worldsheet, we find semiclassically
in Euclidean space that this causes the deformation of
closed string diagrams to be localized near the boundary of AdS
space.  In Lorentzian space we do not expect this to
be the case, and we present some indirect evidence
(coming from the behavior of amplitudes in the flat space limit)
that in Lorentzian space closed string amplitudes are affected in the bulk.

We also study explicitly the dynamics of D-branes.
Diagrams involving D-branes have explicit bulk effects
which are evident semiclassically in Euclidean space,
and we explicitly compute the contribution of
the deformation to bulk forces between D-branes.

We also discuss the deformation
in the language of the low energy effective theory.
The deformation we perform is by a product of currents,
each of which is dual to a gauge field in the bulk with
a Chern-Simons coupling at leading order in the low-energy expansion
(see, for instance, \BalasubramanianRT).
The deformation of the dual CFT action
by a product of chiral and antichiral currents can
be identified with a local deformation of the boundary (surface) terms in
the gravity-side $2+1$-dimensional Chern-Simons theory in a standard way
\refs{\WittenHF,\ElitzurNR,\MaldacenaSS}.  This description is equivalent
in this case to our description \genwsdef\ (both descriptions
lead to the same perturbation expansion involving
insertions of bulk-boundary propagators),
and leads equivalently
to interesting bulk physics such as novel contributions to forces between
D-branes.  It is also worth emphasizing that even though the surface term
is local in the $3d$ action on $AdS_3$, it is non-local in the $6d$ action on
$AdS_3\times S^3$, with a non-locality scale given by the $AdS$
curvature radius.
We will mostly use the formalism \genwsdef\ which
generalizes to other cases of NLSTs and ``double-trace'' deformations.
It is interesting that in this simple case the NLST results
obtained from a non-local shift in the worldsheet action
can be reproduced by a change in the $3d$ local action involving
boundary terms in spacetime.

The construction of a stable non-supersymmetric background in
perturbative string theory (with flat moduli and maximal symmetry in
the noncompact dimensions) provides one potential application of these
theories. More generally, it is important to articulate the conditions
for consistency of this type of theory directly in string theory
language, in order to understand whether this phenomenon goes beyond
the fascinating but somewhat esoteric realm of AdS spacetimes. In this
work, we find that a particular scaling of the deformation leaves
interesting effects in the flat space limit. It is not clear
if this limit defines a consistent theory or not, but if it does then
this may provide an avenue towards understanding more general
realizations of NLST's\foot{In a companion project \usnext, we
are investigating the role of NLST's in describing squeezed
states, such as those that occur in particle production processes
in time dependent backgrounds, in perturbative string theory.}.

The $3d$ boundary term which generates our deformation
affects the bulk in AdS in two ways.
One has to do with the analogy between AdS and a finite box -- it takes
some modes a finite time to reach the boundary.  Another way
in which the boundary can affect the bulk is via
the fact that the boundary deformation existed for an infinite
time in the past.  The latter effect survives in the flat limit,
along with severe non-locality felt by modes with momentum
along the dimensions descending from the $S^3$.

This paper is organized as follows.  In \S2, we introduce the
basic deformation on the field theory side and then translate it
to the gravity side using the vertex operators of \KutasovXU.  In
\S3, we study the effects of the deformation on closed string
correlators. In \S4 the description of the deformation in the
low-energy effective theory in three dimensions is discussed.
As mentioned above, this is simply given by a local boundary term
in this case.
Then, in
\S5, we calculate corrections to forces between D-branes (and to
the instanton action of D-instantons) induced by the NLST
deformation. Finally, in \S6 we exhibit a scaling of the
deformation parameter in which these effects survive in the flat
space limit.

\newsec{The Deformation}

In this section we introduce the ``double-trace'' deformation we are
turning on and calculate its effects on correlators on the CFT
side.  We then translate the deformation to the gravity side
language using the vertex operators of \KutasovXU.  In the subsequent
sections
we will calculate the effects of the deformation on physical
quantities directly on the gravity side.

\subsec{Field Theory Side}

Consider an $AdS_3$ background of superstring theory which is dual
to a two dimensional (super-)conformal CFT containing holomorphic and
antiholomorphic $U(1)$ affine Lie algebras of level $k$ generated
by currents $J(x)$ and $\tilde J(\bar x)$ (obeying
$J(x) J(0) \sim k / x^2$). For example, in cases
where the dual CFT has $\cn=(4,4)$ supersymmetry, there is an
$SU(2)\times SU(2)$ R-symmetry and we will be interested in a
$U(1)\times U(1)$ subgroup of this. The dual CFT could also
include sigma-models on circles (there are 8 such circles in the
CFT which is dual to string theory on $AdS_3\times S^3\times T^4$,
which is related by marginal deformations to the sigma-model on
$[(T^4)^N/S_N \times T^4]$ \LarsenUK),
in which case we can choose $J$ and $\tilde J$ to be the
generators of the corresponding isometries.

Our main interest is in the deformation of the dual CFT by
\eqn\basicdef{ \delta S_{CFT}=h \int d^2x J(x)\tilde J(\bar x), }
where $h$ will be normalized shortly.  This deformation is exactly
marginal (as can be seen for example by bosonizing the currents).
In the case that $J$ and $\tilde J$ are part of the R-symmetry
group of a superconformal theory, this deformation completely
breaks the supersymmetry.  This combination of exact marginality
and SUSY breaking is very interesting, as it means for example
that no destabilizing potential for moduli is generated in the
dual string theory at all orders and nonperturbatively.

Many aspects of the effect of the deformation on the dual CFT can
be calculated exactly, since the currents involved in the deformation
\basicdef\ can be bosonized.
It will be convenient to use such a bosonized description,
in which we identify $J(x) = \sqrt{2k} \del_x \eta(x,\bar x)$ and
$\tilde J(\bar x) = \sqrt{2k} \del_{\bar x} \tilde\eta(x,\bar x)$,
where $\eta$ and $\tilde\eta$ are canonically normalized scalar fields.

In the case of the CFT dual to the near horizon limit,
$AdS_3\times S^3\times T^4$, of $Q_1$ fundamental strings and $Q_5$
NS5-branes on a $T^4$, the parameters of the CFT and
those of the background are related as follows\foot{In this case it
was argued in \LarsenUK\ that the CFT which is dual to the
perturbative string theory actually includes some specific terms of
the form \basicdef. So, in this case our discussion will refer to
adding additional terms of this type beyond the terms which are
already present in the ``standard'' string theory.}. The central charge
of the dual ${\cal N}=(4,4)$ SCFT
is $c=6Q_1Q_5$ (up to a correction of order one which
we will ignore, since we will be interested in the perturbative
weakly-curved limit
of $Q_1 \gg Q_5 \gg 1$), and the level of its $SU(2)$
affine Lie algebra is $k=2Q_1Q_5$.  The gravity side AdS radius in
string units is $\sqrt{Q_5}$, and the six-dimensional string
coupling on $AdS_3\times S^3$ is $g_6=\sqrt{Q_5/Q_1}$.  Therefore
powers of $g_6$ correspond to powers of $1/\sqrt{Q_1}$; this will
be important in comparing gravity side diagrams to the expansion
of correlation functions on the field theory side.


Let us proceed with the analysis for the $U(1)$ currents coming from
the $SU(2)$ R-symmetry, for
definiteness. In this case we have
\eqn\currtwo{J(x)J(0)\sim {{2Q_1Q_5}\over{x^2}}.}
This scales as
$1/g_6^2$, which is appropriate since it is related by the duality to
a classical kinetic term for bulk gauge
fields. In the bosonized language we can write our deformation in this
case as
\eqn\defus{{{\tilde h(Q_5)}\over{Q_1Q_5}}\int d^2x J(x)\tilde
J(\bar x)=4\tilde h(Q_5)\int
d^2x\partial\eta\bar\partial\tilde\eta,}
where we normalized the coefficient using the fact
\AharonyPA\ that the deformation should scale
as $g_s^2$ in order to get a reasonable perturbation expansion\foot{
As just discussed, in $AdS_3\times S^3$ with NS charges the only
place $Q_1$ appears is in the string coupling, so counting
powers of $Q_1$ is the same as counting powers of $g_6$.},
and we defined $h \equiv {\tilde h} / Q_1 Q_5$, where apriori
$\tilde h$ can have an arbitrary dependence on $Q_5\sim
L_{AdS}^2/l_s^2$. This normalization is natural from the dual CFT
point of view, since at a generic point of the field theory
moduli space $Q_5/Q_1$ plays no special role, but the central charge
is always proportional to $Q_1 Q_5$. On the string theory side a more
natural choice might be $h \equiv \th' g_6^2 = \th' Q_5 / Q_1$ which
differs from the choice above by $Q_5^2$; we will see that indeed this
choice will be more natural when we discuss the flat space limit in \S6.


The operators of the dual CFT are of the form
\eqn\genop{ {\cal O}_I=e^{i(p_I\eta+\tilde
p_I\tilde\eta)}P_I(\partial^n\eta, \bar\partial^{\tilde n} \tilde\eta) \hat
O_I, }
where $P_I(\partial^n\eta,\bar\partial^{\tilde n} \tilde\eta)$
denotes a polynomial in arbitrary
derivatives of $\eta,\tilde\eta$, and where $\hat O_I$ is an operator in the
coset obtained after dividing by the $U(1)\times U(1)$ bosonized
by $\eta,\tilde\eta$. It is important to emphasize that there is a
particular correlation between the coset part $\hat O_I$ and the free
part $e^{i(p_I\eta+\tilde p_I\tilde\eta)}P_I(\partial^n\eta,
\bar\partial^{\tilde n} \tilde\eta)$ encoded in the set of operators which exist in
the CFT.  In our main example where $J$ and $\tilde J$ are part of
the R-symmetry of the dual CFT, different components of the
spacetime supermultiplets in the undeformed theory have different
R-charges $q, \tilde q$.  The deformation \defus\ breaks
supersymmetry as it couples to these different components according
to their charges.  These R-charges are $SU(2)$ charges:
we thus have
$J(x) e^{ip\eta}(0) \sim qe^{ip\eta}(0)/x$ where $q$ is the
$SU(2)$ weight (integer or half integer) of the operator.
This means that the
charges $p, \tilde p$ which exist in the theory scale as
\eqn\qscale{p\sim q/\sqrt{4Q_1Q_5}, ~~~~ \tilde p\sim {\tilde
q}/\sqrt{4Q_1Q_5}.}

The simplicity of our deformation \basicdef\ allows us to
determine explicitly the effect of the deformation on correlation
functions of the ${\cal O}_I$, starting from the basic Ward
identities
\eqn\ward{\eqalign{&J(x)J(0)\sim {2Q_1Q_5\over x^2},\cr &
J(x)e^{ip\eta}(0)\sim {{\sqrt{4Q_1Q_5}p}\over x}e^{ip\eta}(0).\cr
} }
%


\fig{The leading contribution, at order
$\tilde h g_6^0$,  to the renormalization
of the dimension of charged operators $Y_{\pm q,\pm\tilde q}$ (denoted by
straight lines) by the ``double-trace''
deformation $J\tilde J$ (denoted by the slashed lines meeting at
a boundary point $x$).}{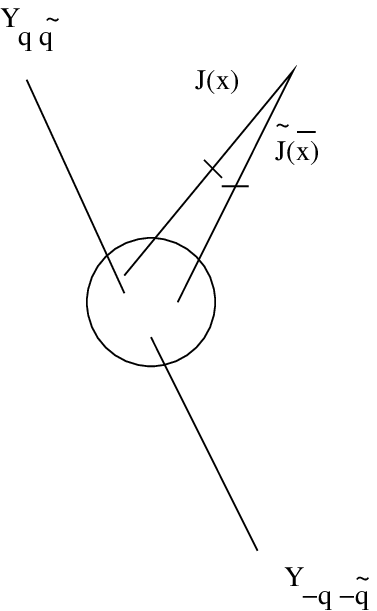}{4 truecm}
\figlabel\dimcorr

One basic effect of the deformation is a shift in the dimension of
charged operators of the form
$Y_{p,\tilde p}\equiv \sqrt{Q_1}e^{i(p\eta+\tilde p\tilde\eta)}$ (for
which we chose an arbitrary normalization such that the 2-point
function scales as $1/g_s^2$). A simple computation gives
\eqn\Ydim{\eqalign{ \delta_{\tilde h}\langle Y_{p,\tilde
p}(x,& \bar x) Y_{-p,-\tilde p}(0)\rangle =\cr & Q_1
\sum_{n=1}^\infty {(4\th)^n\over {n!}}
\int \prod_{i=1}^n d^2x_i\langle
e^{ip\eta}(x)\prod_{i=1}^n\partial\eta(x_i) e^{-ip\eta}(0)\rangle
\langle e^{i\tilde p\tilde \eta}(\bar
x)\prod_{i=1}^n\bar\partial\tilde\eta(\bar x_i)e^{-i\tilde
p\tilde\eta}(0)\rangle.\cr}}
This expression is a power series in the ``double-trace'' coefficient
$\tilde h$ and in the string coupling $g_6^2\sim 1/Q_1$  (the
latter statement follows from the form of \Ydim\ combined with the
scaling \qscale\ of the charges).
The corresponding diagrams on the gravity
side are of effective genus $\ge 1$, with the first
contribution arising at ${\cal O}(\tilde h g_6^0)$ as depicted
in figure \dimcorr.

Let us
evaluate this explicitly at order $\tilde h$. Working out the
correlators this reduces to
\eqn\Ynext{{{4Q_1\tilde h p\tilde p}\over{x^{p^2/2}\bar x^{{\tilde
p}^2/2}}}\int d^2 x_1\biggl| {1\over{x_1-x}}-{1\over x_1}\biggr|^2.}
This integral is logarithmically divergent when $x_1$ approaches
the other operators $Y$ at $x$ and at 0 (the log divergence for large
$x_1$ cancels among the different terms in
\Ynext). Let us include a UV cutoff $a$, which cuts off the integrals
such that for any other operator insertion at $x_0$, the
range of $x_1$ is bounded by $|x_1-x_0|\ge a$. Doing the integral
in \Ynext, one then finds
\eqn\Yfin{ \delta_{\tilde h}\langle Y_{p,\tilde p}(x,\bar x)
Y_{-p,-\tilde p}(0)\rangle=8\pi Q_1\tilde h{{p\tilde
p}\over{x^{p^2/2}\bar x^{{\tilde p}^2/2}}}\log{{{|x|^2}\over{|a|^2}}}.}
The $\log|a|^2$ piece must be absorbed in a redefinition of the
operators $Y_{p,\tilde p}$ as is standard in conformal
perturbation theory \DijkgraafJT\ (see also the discussion of this in
\S3.1). Namely, here
\eqn\Yshift{ Y_{p,\tilde p}\to Y_{p,\tilde p}+(8\pi \tilde h ~
p\tilde p ~ \log ~ a ) ~ Y_{p,\tilde p}.}
What remains amounts to a shift in the dimension of $Y$ by
\eqn\YdimII{(-8\pi\tilde h p\tilde p,-8\pi\tilde h p\tilde p)
}
to first order
in $\tilde h$. Taking into account the scaling \qscale\ of the
charges, this shift is of order $\th g_6^2$ (for small charges).
It is easy to generalize this to general correlation functions.


One can similarly work out changes to correlators involving
currents (and their descendants) arising from our deformation.
For example,
\eqn\defJ{ \delta_{\tilde h}\langle J(x)J(0)\rangle =
\sum_{n=1}^\infty \biggl({{\tilde
h}\over{Q_1Q_5}}\biggr)^n{1\over{n!}}\int\prod_{i=1}^n d^2x_i\langle
J(x)\prod_{i=1}^n J(x_i) J(0)\rangle\langle\prod_{i=1}^n\tilde
J(\bar x_i)\rangle. }
Here only even $n$ contributions survive.  All these contributions are
(since they involve $n+1$ contractions of $J$'s)
at order $Q_1\sim g_s^{-2}$, the same order as tree-level diagrams.
This agrees with the set of diagrams that
contribute to \defJ\ on the gravity side, which involve $n+1$
disconnected spheres (connected by insertions of the deformation).
The first contribution,
at order $\tilde h^2$, is given by
\eqn\JJint{4 Q_1Q_5\tilde h^2
\int d^2x_1 d^2x_2{1\over{(\bar x_1-\bar
x_2)^2}}\biggl[{1\over{(x-x_1)^2}}{1\over
x_2^2}+{1\over{(x-x_2)^2}}{1\over
x_1^2}+{1\over{(x_2-x_1)^2}}{1\over x^2}\biggr].}
The last term here is related to a divergence in the vacuum
amplitude,
\eqn\dvacuum{\eqalign{\delta_{\th}\vev{1} & = \sum_{n=1}^{\infty}
\biggl({{\tilde
h}\over{Q_1Q_5}}\biggr)^n{1\over{n!}}\int\prod_{i=1}^n d^2x_i\langle
\prod_{i=1}^n J(x_i) \rangle\langle\prod_{i=1}^n\tilde
J(\bar x_i)\rangle = \cr
& = 2 \th^2 \int d^2x_1 d^2x_2 {1\over {|x_1-x_2|^2}} + \cdots, \cr }}
so it will cancel when we compute the properly normalized correlation
function which involves dividing by $\vev{1}$.

The first two terms in \JJint\ give identical finite results, adding up to
\eqn\JJshift{\eqalign{
2\cdot 4Q_1Q_5\tilde h^2\int d^2x_1d^2x_2 & {1\over{(\bar x_1-\bar x_2)^2}}
{1\over{(x-x_2)^2}}{1\over x_1^2}=\cr
& = 8Q_1Q_5\tilde h^2\int d^2x_1d^2x_2
{\del\over{\del\bar x_2}}({1\over{\bar x_1-\bar x_2}})
{\del\over{\del x_2}}({1\over{x-x_2}}){1\over x_1^2}=\cr
& = 8 Q_1Q_5\tilde h^2\int d^2x_1d^2x_2
{\del\over{\del x_2}}({1\over{\bar x_1-\bar x_2}})
{\del\over{\del \bar x_2}}({1\over{x-x_2}}){1\over x_1^2}=\cr
& = 32 \pi^2 Q_1Q_5\tilde h^2\int d^2x_1d^2x_2 \delta^{(2)}(x_1-x_2)
\delta^{(2)}(x-x_2){1\over x_1^2}=\cr
& = {{32\pi^2 Q_1Q_5\tilde h^2}\over x^2}.\cr
}}
If desired, one can always renormalize $J$ by a multiplicative constant
(depending on $\th$) which will cancel this correction and keep the
same form of $\vev{J(x) J(0)}$.

Another example is
\eqn\deftJ{\delta_{\tilde h}\langle J(x)\tilde J(0)\rangle =
\sum_{n=1}^\infty \biggl({{\tilde
h}\over{Q_1Q_5}}\biggr)^n{1\over{n!}}\int\prod_{i=1}^n d^2x_i\langle
J(x)\prod_{i=1}^n J(x_i)\rangle\langle\tilde
J(0)\prod_{i=1}^n\tilde J(\bar x_i)\rangle. }
Here only odd values of $n$ contribute.  For $n=1$, this is
\eqn\JtJint{\eqalign{4Q_1Q_5 \tilde h\int
d^2x_1{1\over{(x-x_1)^2}}{1\over{\bar x_1^2}} &=
-4Q_1Q_5\th \int d^2x_1 {{\del}\over{\del x_1}} ({1\over {x-x_1}})
{{\del}\over{\del {\bar x}_1}} ({1\over {{\bar x}_1}}) = \cr
&= -4Q_1Q_5\th \int d^2x_1 {{\del}\over {\del {{\bar x}_1}}} ({1\over
{x-x_1}}) {{\del}\over {\del x_1}} ({1\over {{\bar x}_1}}) = \cr
&= 16\pi^2 Q_1Q_5\th \int d^2x_1 \delta^{(2)}(x-x_1) \delta^{(2)}(x) =\cr&=
16\pi^2 Q_1 Q_5 \th \delta^{(2)}(x), \cr }}
which is just a shift in the contact term between $J$ and $\tJ$.
We can swallow this by redefining the original contact term (the
same will be true at higher orders as well).


By using exact formulas for correlators involving $\eta$
and $\tilde\eta$, we can
in principle calculate explicitly the effects of the deformation
on all operators \genop\ of the theory, including the parts
involving complicated descendants.  It is worth emphasizing,
however, that the set of operators \genop\ has a lot of structure.
The AdS/CFT correspondence maps all states in global AdS to operators
in the CFT, so operators of this form
describe all possible bulk excitations on the gravity
side. The CFT charge $q$ maps to the charge under the corresponding gauge
field on $AdS_3$ (given by the integral of the gauge field around the
boundary of $AdS_3$ at fixed time in global coordinates). Clearly,
there are many configurations with total
charges $q,\tilde q$; the information about the distribution of
this charge in the bulk of the spacetime is encoded in the details
of the $P\hat O$ factors in the operator.
It is interesting
that the formula \YdimII\ implies that the change in the dimension of
operators (and, therefore, the change in the energy of the corresponding
states in global AdS) depends only on their charge.
However, in order to understand
the effects of our deformation on the dynamics of nontrivial
distributions of charge in the bulk of the space, one needs to
keep track of the ``fine structure" in the operators.

In particular, in \S5, we will be interested in forces between
separated D-branes in the bulk of $AdS_3\times S^3$.  Pairs of
$D0$-branes in the bulk of $AdS_3$ are not quite in stationary
states, as there are forces between them (which are small for
large $L_{AdS}$).  Such a pair is therefore described by a
combination of operators \genop\ which does not form an eigenstate
of the dilatation operator in the dual CFT.  This can be modeled
by a sum of an operator of particular dimension plus $1/L_{AdS}$
times an operator or sum of operators of different dimension.
After the deformation, the correlation functions of the different
terms scale in different ways determined by their correlators with
$J,\tilde J$ as in the simple examples worked out above.  The
force term is still multiplied by a small coefficient,
$1/L_{AdS}$, but its magnitude will in general receive
corrections.  We will calculate this effect explicitly for some
D-branes in \S5, and reproduce this general structure predicted by
the dual CFT.

\subsec{The Gravity Side}

The general
formalism described in \AharonyPA\ implies that deforming the CFT
by a ``double-trace'' operator of the form
$h \int d^2 x {\cal O}_1(x) {\cal O}_2(x)$ is described in
string theory, at least to leading order in $h$, by deforming the
worldsheet action by the non-local term $h \int d^2x \int d^2 z_1
V_1(z_1;x) \int d^2 z_2 V_2(z_2;x)$, where $V_{1,2}(z;x)$ are the
vertex operators for ${\cal O}_{1,2}(x)$. In our case, as
described in \KutasovXU, the affine Lie algebra generated by $J(x)$ in
the dual CFT is related to an affine Lie algebra generated
by $k(z)$ on the worldsheet. An insertion of $J(x)$ into a CFT
correlation function is equivalent to an insertion of $K(x)$
defined by \eqn\Jvertex{K(x) = -{1\over \pi} \int d^2z k(z)
\del_{\bar z} \Lambda(z,{\bar z};x, \bar x)} in the string
worldsheet, where $\Lambda$ is a particular operator such that
$\del_{\bar z} \Lambda(z,\bar z;x, \bar x)$ is a primary operator
of the worldsheet conformal algebra with dimension $(0,1)$, and
also a primary of the space-time conformal algebra with scaling
dimension $(1,0)$. We wrote down the vertex operator for the
bosonic string; in the case of the superstring (which is the
case we are interested in) there will be some additional terms
in the expression above, but they do not change our discussion
and our semi-classical computations below so we will not write
them down explicitly.

If we choose coordinates on $AdS_3$ such that
the string-frame metric is of the form
$ds^2 = Q_5 (d\phi^2 + e^{2\phi} d\gamma d{\bar
\gamma})$ (where the curvature in string units is $-2/Q_5$), then
we can write an expression for $\Lambda$ in terms of the worldsheet
fields $\phi(z,\bar z), \gamma(z,\bar z)$ and $\bar \gamma(z,\bar z)$,
in the semi-classical approximation, of
the form
\eqn\lambdavert{\Lambda(z,\bar z;x, \bar x) =
-{{(\bar \gamma - \bar x) e^{2\phi}}\over {1 + |\gamma - x|^2
e^{2\phi}}}.}

The deformation of the worldsheet Lagrangian corresponding to
\defus\ is given by
\eqn\wsdef{\delta S_{worldsheet} = {\th\over
{Q_1 Q_5 \pi^2}} \int d^2 x \int d^2 z_1 \int d^2 z_2 k(z_1) \del_{\bar z_1}
\Lambda(z_1,\bar z_1;x, \bar x) {\tilde k}(\bar z_2) \del_{z_2}
{\bar \Lambda}(z_2, \bar z_2;x, \bar x).}
The vertices \Jvertex\
have many interesting properties that were analyzed in \KutasovXU\
and used there to derive the Ward identities for the current
$J(x)$. Since $\del_{\bar z} k(z) = 0$ except for delta function
contributions at the locations of other vertices, we can integrate
by parts and write \Jvertex\ as a contour integral of $k \Lambda$
on contours surrounding the insertion points of vertex operators,
and (if they exist) on boundaries of the worldsheet (note that there
are no singularities when the vertex operators in $K(x)$ and $\tilde K(
\bar x)$ approach each other). In
particular, the vertex operator $K(x)$ \Jvertex\ can be written
in the form
\eqn\contform{ K(x)=\sum_{insertions, boundaries}\oint
{dz\over{2\pi i}}k(z)\Lambda(z,\bar z; x,\bar x).}
 This leads \KutasovXU\ to the Ward identity for correlators
of $K$ with charged fields.  Let $W_q(x)$ be the integrated vertex
operator corresponding to a
primary of the $J$ affine Lie algebra with charge $q$, so that
correspondingly it is a primary
of the corresponding worldsheet affine Lie algebra with charge
$q$. Then, one finds \KutasovXU\
\eqn\wardid{ \vev{K(x) \prod_i W_{q_i}(x_i,\bar x_i)}=
\sum_i{{q_i\over{x-x_i}}}\vev{\prod_i W_{q_i}(x_i,\bar x_i)} }
for closed string
worldsheet correlation functions, reproducing the Ward identities
of the dual CFT.
Many interesting operators (including $J(x)$ itself) will not
have this property of being primaries of charge
$q$ and then we will have more complicated
expressions for their correlation functions, as discussed in \S2.1.

\newsec{Effect of the Deformation on Closed String Amplitudes}

Now, let us take some correlation function of closed string vertex
operators in the theory before the deformation, and consider the
effect of the deformation on the correlation function. In
perturbation theory the effect of the deformation is given by
the insertion of some number of $K(x_i)$ and $\tilde K(\bar x_i)$
vertex operators into the correlation function, and integrations
over $x_i$. If the correlation function involves only primary
fields we can then easily compute it on the worldsheet using
\wardid, and it is obvious that we reproduce the CFT computations
of the same correlation functions \Ydim -\JtJint\
described in \S2.1.

Our deformation is exactly marginal and affects physics
at all scales on the field theory side, and we
have introduced various changes to
correlation functions of closed strings in $AdS_3$, so we might
expect the bulk physics to be affected by the deformation,
and perhaps to become non-local (with a non-locality scale much
bigger than the string scale). For
the case of a double-trace deformation in $AdS_5$ various arguments
for bulk non-locality were given in \AharonyPA. However, in our case
we need to be more careful because, as discussed above, the vertex
operators we deform by are total derivatives on the worldsheet, so it
is not clear that the deformation is really felt all over the worldsheet.
Semiclassical worldsheets in Euclidean $AdS_3$
stretch all the way to the boundary, where the vertex operators describing
external states in the Feynman diagrams are inserted \deBoerPP.
It is straightforward to check, using the methods of \deBoerPP,
that the insertion of $K(x)$ does not change the shape of the
saddle point configuration of the worldsheet near the vertex
operator insertions at the boundary.
The worldsheet path integral of course involves integration over
all worldsheet shapes, but from \deBoerPP\ we see that the dominant
(saddle point)
contribution is one in which the $W_q$ insertions are
at the boundary. As discussed above, further insertions of $K(x)$
localize at the same points on the worldsheet.
Thus, in this special case where the vertex operators
we deform by are total derivatives, it seems that the only effect
evident semiclassically on Euclidean
closed-string amplitudes is localized at the boundary of $AdS$ space.

The case of more physical interest on the gravity side is
the Lorentzian case, where scattering events can take place
in the bulk of the space.  For the Lorentzian case we will provide an
indirect argument in \S6, based on features
of the flat space limit,
that the effects of our deformation are felt also in the bulk
of the space and not just near the boundary.

The existence of
non-supersymmetric shifts of charged closed string masses obtained from the
shifted dimensions \YdimII, combined with the exact stability of
the model, raises the fascinating question of how to see the
cancellation of the moduli potential directly on the gravity
side of the correspondence.  We will return to this question in \S3.2 after
considering the divergence structure of the deformation on the
gravity side.

\subsec{Regularization of Divergences}

In studying marginal deformations of CFTs in
conformal perturbation theory, one encounters divergences in
calculating corrections to correlation functions, which can
be consistently regularized and absorbed in rescalings of
the operators (see e.g. \DijkgraafJT).  The cutoff $a$
we introduced in
\Yfin\ and the rescaling \Yshift\ are an example
of this procedure in our case on the field theory side.
We would now like to illustrate how this regularization
is described on the gravity side. This can be deduced
by using the UV/IR correspondence.

On the gravity side, the first-order correction in a correlator like
\Ydim\ is of the form
\eqn\gravdim{ {\tilde h \over {Q_1 Q_5}}
\int d^2x_1\langle W_{q,\tilde q}(x,\bar x)
W_{-q,-\tilde q}(0)K(x_1)\tilde K(\bar x_1)\rangle. }
Anticipating
that the result will be divergent, let us put an IR cutoff in
space-time at a finite value of $\phi$, leaving the region
$\phi < \phi_c$, and use the
semiclassical analysis of the worldsheet and of $\Lambda$.
Taking into account the localization of $K$ at the $W$ insertions
\contform\ and the fact that $\oint k(z) {dz \over {2\pi i}}$ measures
the charge, this becomes
\eqn\gravnext{\delta_{\tilde h}\langle WW\rangle =
{\tilde h \over {Q_1 Q_5}} \int d^2x_1q\tilde q
\biggl|\Lambda_1(x_1)-\Lambda_2(x_1)\biggr|^2 \vev{WW},}
where $\Lambda_1$
and $\Lambda_2$ refer to the semiclassical value of $\Lambda$ at
the positions of the two $W$ insertions (cut off at $\phi_c$).
For large $\phi_c$ we find
\eqn\semiLam{\eqalign{ & \Lambda_1(x_1)= -{{(\bar x - \bar
x_1)}\over{e^{-2\phi_c}+|x-x_1|^2}},\cr & \Lambda_2(x_1)= -{{(- \bar
x_1)}\over{e^{-2\phi_c}+|x_1|^2}},\cr } }
where we have replaced
the $\gamma$ coordinate of each insertion by its boundary value
($x$ or 0 respectively) since the corrections to this
value are subleading at large $\phi_c$
to the $e^{-2\phi_c}$ contribution
we have included.
Plugging \semiLam\ into \gravnext\ gives an $x$ integral whose
log divergence at large $x_1$ cancels among the various terms in
\gravnext\ (just like in \Ynext). The leading divergent
behavior when $x_1$ approaches the other
insertions at $x$ and $0$, and as $\phi_c \to \infty$, is
\eqn\cutdiv{
\int d^2 w{{|w|^2}\over{(e^{-2\phi_c}+|w|^2)^2}}
\sim -2\pi \log(e^{-2\phi_c})=4\pi\phi_c.
}
Now that we have expressed the cutoff divergence in terms of
gravity side quantities, we can absorb this divergence
into a rescaling of the vertex operators $W_{q,\tilde q}$,
corresponding to the rescaling \Yshift\ we had on the
field theory side.
In string theory, we can further translate this
cutoff into a short-distance cutoff on the worldsheet
using \deBoerPP.
The IR cutoff $\phi_c$ in the target space geometry
corresponds to a cutoff
\eqn\wsa{
a_{worldsheet}(h)=e^{-{\phi_c\over{4 h}}}
}
on the worldsheet near an insertion of a vertex
operator corresponding to a scalar operator
of dimension $h(=\bar h)$ in the dual CFT.

\fig{Vacuum diagram at order $\tilde h^2 g_s^0$.  The
insertions of the vertex operators in the ``double-trace''
deformation are indicated by the pair of lines with slashes joined at
the boundary.}{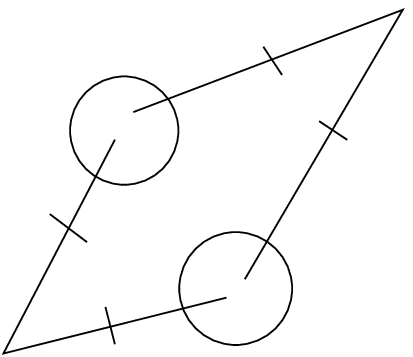}{3 truecm}
\figlabel\vacfig

\fig{Modulus tadpole at order $\tilde h^2 g_s^0$.  The
insertion of the vertex operator for the modulus field is
indicated by the plain line.}{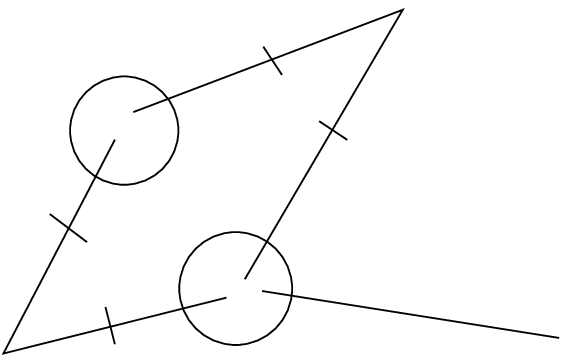}{3 truecm}
\figlabel\modfig

There are also formal divergences in contributions to
the vacuum amplitude in the bulk.  For example, the diagram
in figure \vacfig\ has a logarithmic divergence (given by \dvacuum).
These diagrams by
themselves are not physically observable -- they map to $\vev{1}$ in
the CFT which we should always choose to equal one. However, the ratio
between any other diagram and the sum of vacuum diagrams is observable.
For example, we can look at the
same diagram probed by an external line as depicted in
figure \modfig. This will be relevant for the moduli potential, which we
turn to next.

\subsec{The Moduli Potential in String Perturbation Theory}

As discussed above, when we deform the CFT which is dual to string
theory on (say) $AdS_3\times S^3\times T^4$ by a deformation
\basicdef\ involving $U(1)_R$ currents, we explicitly break the
space-time supersymmetry. From the space-time point of view we would
naively expect to generate a moduli potential in such a case, such that
not every point in the original moduli space would still give a stable
background after the supersymmetry breaking. However, we
know that this does not happen in our case since the deformation in the CFT is
exactly marginal (independently of any of the other parameters of the
CFT), so we expect to have an exact non-supersymmetric background after
the deformation with the isometries of $AdS_3$ for any value of the
other moduli of the theory.
We are using a slight abuse of terminology here:  since
in general NLSTs do not have a local effective action, the notion
of a {\it moduli potential} may not persist. However, we can still ask
whether all the moduli of the original theory remain, and do not develop
tadpoles even after we add the supersymmetry breaking deformation. We
have moduli operators ${\cal O}_{modulus}^{(I)}(x,\bar x)$
which are of dimension $(1,1)$, and the
vanishing of a term of order $m$ in the fields in
the original ``moduli potential'' is manifested in the vanishing of
the integrated
correlation function of $m$ of these operators in the CFT\foot{The
case $m=2$ actually does not vanish; it is related to
the propagator on the gravity side, and diverges after we integrate
over $x$. The vanishing of the quadratic term in the
``moduli potential'' is accounted for by the dimension of the modulus
operator, which corresponds to a massless field on the gravity
side.}.
 From the dual CFT it is clear that this must still be the case also
after the deformation, and in
this section we will see how this happens from the point of view of
string perturbation theory in the bulk (which gives part of the
contribution to the correlation functions in the full dual CFT).

In usual flat-space string theory, when we break supersymmetry we
would expect to have a non-zero torus vacuum amplitude. There, this
amplitude
is proportional to the torus diagram with an insertion
of the zero momentum dilaton, which is the worldsheet manifestation
of the fact that the vacuum energy in perturbative string theory
is really a potential for the dilaton.  Our situation is different
since the dilaton is a fixed scalar and therefore massive. Thus, we
would expect to generate a potential only for the other moduli which
actually correspond to massless fields on $AdS_3$.
In any case the vacuum diagram by itself has no physical meaning, so
we cannot use it to learn about supersymmetry breaking in the bulk;
the physical effects of the vacuum energy are encoded in the diagrams
with an external graviton or moduli line, which determine the
curvature and moduli dynamics generated by the vacuum energy.

In the case we are interested in here, the moduli involve the $T^4$ part
of the worldsheet CFT; for most of the moduli the vertex operator
corresponding to
$\int d^2x {\cal O}_{modulus}(x,\bar x)$ is simply $\int d^2z \del X^i \bar\del
X^j$ (the others come from the RR sector and our argument in the next
paragraph will apply to them as well).
The leading correction to the moduli tadpole after the deformation
comes from figure \modfig. It is easy to see that this vanishes,
because the worldsheet
correlation function on one of the spheres factorizes into
a correlation function involving the $T^4$ directions and one
involving the $AdS_3\times S^3$ directions.  The first factor
is just of the form $\langle :\partial X^i\bar\partial X^j:\rangle $ where
the $X^i$ are embedding coordinates of the string in the $T^4$
directions.  This vanishes.

Next, let us consider arbitrary diagrams contributing to the ``moduli
potential'', at a
general order in the perturbation theory in $g_s$ and $\th$.
Such a digram would have various connected components, which are genus
$g$ surfaces
with some number $n$ of insertions of $J$, $\tilde n$
insertions of $\tilde J$, and $m$ insertions of $\int d^2x {\cal
O}_{modulus}^{(I)}(x,\bar x)$ (where $I$ labels the various moduli fields).
This subdiagram is a correlator in the original undeformed theory, of
the form
\eqn\genmodcorr{\langle J(x_1)\dots J(x_n) \tilde
J(\bar x_{n+1})\dots \tilde J (\bar x_{n+\tilde n}) \int d^2x {\cal
O}_{modulus}^{(1)}\dots \int d^2x {\cal O}_{modulus}^{(m)}\rangle_{genus\ g}.}
If $n=\tilde n=0$, the diagram is identical to a contribution to the
``moduli potential'' in the undeformed supersymmetric theory, which
cancels\foot{More precisely, this subdiagram is a particular term in the
expansion of the CFT ``moduli potential'' in powers of $g_s^2 = Q_5 /
Q_1$, but since the full correlation function vanishes every term in
its expansion must vanish as well.}.
For the other diagrams which feel the deformation and therefore
the supersymmetry breaking, we note that the moduli of the torus
(which are the scalar fields on $AdS_3$ we are discussing here) are
uncharged under the $U(1)$ isometries generated by $J$ and $\tilde J$,
and have a non-singular OPE with the current operators.
As discussed above, the vertex operators for $J$ and $\tilde J$ are
total derivatives on the worldsheet which can be written as integrals
around the other insertion points, and (as in \KutasovXU) these
integrals get no contributions near the moduli operators. Thus, ignoring
 the picture changing operators inserted
on the Riemann surface at higher genus, which
include terms from all sectors of the worldsheet CFT
and can lead to additional singularities,
one would find
that the
correlation function \genmodcorr\ factorizes into the part involving
$J$ and $\tilde J$ times the part involving the moduli, and the latter
vanishes as argued above.  This calculation
of the $n+\tilde n+m$-point function can be done equivalently in
the dual CFT description of
the original theory, where it cancels by an exact factorization
argument, and one therefore deduces that the full calculation of the diagram
including the picture changing operators still leads to a cancellation.
Thus
we see also on the string theory side
that we do not produce a ``moduli potential'', despite the absence of
supersymmetry.

One might worry that there could be moduli which have a singular OPE
with the currents $J$ or $\tilde J$. If we bosonize the currents as in
section 2, then because the ${\cal
O}_{modulus}^{(I)}$ are dimension $(1,1)$ operators in the
dual CFT and they are uncharged under $J, \tilde J$, they could
only depend on $\eta,\tilde\eta$ by a factor of $\del\eta$ or
$\bdel\tilde\eta$. So, we can write these operators generally as
${\cal O}_{modulus}^{(I)} = {\cal O}_0 + \del\eta(x) \hat{\cal
O}_R(\bar x) + \hat{\cal O}_L(x) \bdel\tilde\eta(\bar x)$ where ${\cal
O}_0$ has a non-singular OPE with the currents, $\hat{\cal O}_R$ is a
dimension $(0,1)$ operator and $\hat{\cal O}_L$ is a dimension $(1,0)$
operator. Note that the last two terms are actually
``double-trace" operators, since $\del \eta$ is simply
proportional to $J$, and they do not correspond to scalar fields on
$AdS_3$. However, even for moduli of this ``double-trace'' form we can
argue that no tadpoles are generated after our deformation. The same
arguments above show that
the effect of the deformation on the tadpole for these operators
must be proportional to the value of $\langle \hat{\cal O}_R(\bar x)\rangle$
or $\langle \hat{\cal O}_L(x)\rangle$ in the original theory, which
obviously vanishes.

We can also give a direct space-time argument for the vanishing of the
``moduli potential'' after the deformation.
On the gravity side, the vanishing of the ``moduli potential'' after
our deformation corresponds to the statement that in the original
theory before the deformation, the coupling of
Chern-Simons gauge fields
(which are the fields dual to $J,{\tilde J}$)
to the moduli remains zero quantum mechanically.  This follows
by gauge invariance
from the fact that the pure gauge modes $A=d\Lambda$ (whose
field strength vanishes) do not couple only to each other or
to the uncharged moduli
fields at any order in perturbation theory in the original
background.

In any case, the
result is that in our diagrammatic expansion, in perturbation
theory in $\tilde h$, the diagrams contributing potentially
destabilizing contributions to the ``moduli potential'' cancel
by virtue of the vanishing of corresponding diagrams
in the original theory, which appear as subdiagrams in
the deformed theory.  It would be nice to gain a more intuitive
understanding in the bulk spacetime
of how the loop diagrams involving
closed strings in the bulk, which have bose-fermi splitting (using
\YdimII, since the bosons and fermions have different charges under
$U(1)_R$),
manage to cancel in this
theory.  We will return to this in \S5.2\ after studying some bulk
effects, including supersymmetry breaking effects, of D-branes
in our theory in \S5.

\newsec{Effect of the Deformation on the Low-energy Action}

In \S3 we saw indications that when computing
the $n$-point function in Euclidean space of any set of vertex
operators on the worldsheet, the contribution of the ``double-trace''
deformation is localized at the boundaries of AdS. In this section we
would like to discuss this in the context
of the low energy effective description, and to
clarify from this point of view where boundary terms arise. In the
next section (\S5) we will return to our analysis of the effects of
the deformation in string theory and the stable supersymmetry breaking
mechanism encoded in this model.

In general in a NLST, one would not expect a {\it local} gravity
or supergravity action in the infrared.  In our present case,
which is based on Chern-Simons gauge fields in $2+1$ dimensions,
some simplifications arise if we focus on the $AdS_3$ part
of the geometry\foot{We thank J. Maldacena
for emphasizing this aspect.}.
In particular, from \refs{\WittenHF,\ElitzurNR,\MaldacenaSS}
it follows that if we bosonize the currents as in section 2, then
the bulk Chern-Simons gauge fields which are dual to the CFT operators
$J$ and $\tJ$ are given by
$A=\sqrt{4Q_1 Q_5} d\eta$ and
$\tilde A=\sqrt{4 Q_1 Q_5} d\tilde\eta$ away from sources (where
$\eta$ and $\tilde\eta$ are defined on all of $AdS_3$ and their boundary
value is given by the objects defined in section 2). Then,
one can realize our deformation $4{\tilde h}
\int d^2 x \del\eta {\bar \del}{\tilde \eta}$
by a boundary term in the CS theory
\eqn\Abdry{\delta S_{SUGRA} =
{\tilde h \over {Q_1 Q_5}} \int_{\del AdS_3} A\wedge \tilde A.}

This prescription reproduces our perturbation expansion in
$\tilde h$, as can be seen by regarding \Abdry\ as part of
the interaction Lagrangian in the gravity-side theory.
Bringing down powers of \Abdry\ in the path integral
and contracting the boundary fields $A_\del,\tilde A_\del$ in
\Abdry\ with bulk fields $A_b,\tilde A_b$ coming from
insertions of interaction vertices from the bulk
Lagrangian, one obtains the bulk-boundary propagators implicit
in the vertex operators in \wsdef.  In particular, as we will
see further in \S5, we find significant effects of the deformation
in the bulk arising from this.  These come from the fact that the $AdS_3$
acts like a finite box for some modes, and more generally
from the fact that
the boundary term \Abdry\ is present throughout time.
Note that \Abdry\ is not a local term in six dimensions,
as each of the fields appearing in \Abdry\ is actually in a particular
spherical harmonic on the $S^3$, so writing this term down in the six
dimensional action entails performing two integrations over the
$S^3$. Thus, in the full theory this term is manifestly non-local at
the AdS curvature scale.

In fact, writing the deformation in the form \Abdry\ is a special
case of something we can do in general to
describe deformations in AdS/CFT. Let
us work in Euclidean AdS space with the standard coordinate system $ds^2 =
(dr^2 + dx^{\mu} dx_{\mu}) / r^2$. In conformal
perturbation theory, if we deform the Lagrangian by a ``single-trace''
operator ${\cal O}$ of dimension $\Delta$ which is dual to a SUGRA
field $\phi(x,r)$, $\delta S_{CFT} = h \int d^dx {\cal O}(x)$, then we need
to insert into the dual supergravity picture any number of
boundary-to-bulk propagators of the field $\phi$, each with a
coefficient $h$. One way to do this is to deform the SUGRA action
by a boundary term of the form $\delta S_{SUGRA} = \lim_{r \to 0} h \int d^dx
\phi(x,r) r^{d-\Delta}$, which reproduces the same perturbation
expansion because of the relation between the bulk-to-boundary and
bulk-to-bulk propagators, if we add this term without changing the boundary
conditions on the fields. However, usually this description is not
very useful since the limit $r \to 0$ is singular so we do not get
a local deformation of the action, except in the case $\Delta=d$ of
marginal deformations. For marginal deformations the effect of the
added term at first order in $h$ is simply to change the bulk value of $\phi$
by a constant amount proportional to $h$, as in the usual
description. However, this violates the usual boundary condition for a
massless field (which sets its boundary value to a particular constant),
so this formalism breaks down also in this case (leading to singular
configurations). In any case,
this illustrates that writing the deformation as a local
boundary term does not preclude having large effects of the
deformation in the bulk.

Similarly, also for ``double-trace'' deformations by a product of two
scalar operators, of the form $\delta S_{CFT} = \tilde h
\int d^dx {\cal O}_1(x) {\cal O}_2(x)$, we can reproduce the
perturbation theory in $\tilde h$ by adding to the supergravity action
$\delta S_{SUGRA} = \lim_{r \to 0} \tilde h \int d^dx \phi_1(x,r) \phi_2(x,r)
r^{2d-\Delta_1-\Delta_2}$. Again, this is not very useful since the
added term generally has no good $r \to 0$ limit, and in particular
this happens in the marginal case of $\Delta_1+\Delta_2=d$.
However, if we deform by
vector fields instead of scalar fields, we get a power of
$r^{2d-2-\Delta_1-\Delta_2}$ instead of the power we wrote above. In the
case we are discussing in this paper (for which $d=2,\Delta_1=\Delta_2=1$)
this power vanishes, so we simply
reproduce the deformation \Abdry, which is perfectly well
behaved. Note that, as described for instance in the discussion around
equation (A.19) of \MaldacenaSS, we do not need to impose any
boundary conditions on the fields $A, \tilde A$, since by adding
appropriate boundary terms we can set the relevant currents to be
chiral and anti-chiral by the equations of motion (the Euclidean action is of
the form ${k\over {2\pi}} \int_{AdS_3} (A\wedge dA - {\tilde A}\wedge
d{\tilde A}) - {ik\over {4\pi}} \int_{\del AdS_3} (A\wedge *A +
{\tilde A} \wedge *{\tilde A})$, where the $*$'s are taken in the
boundary of AdS space). Thus, it is not
necessary to change the boundary conditions after deforming by \Abdry,
and this term automatically reproduces the perturbation expansion in
the CFT which we described in section 2.

\newsec{D-branes:  Bulk Effects and SUSY Structure}

In section 3 we studied closed string amplitudes
in which the operators $K(x),\tilde K(\bar x)$ involved in our
deformation localized to the boundary of $AdS_3$ (semiclassically).
When the worldsheet has boundaries on D-branes, $K(x)$
gets additional
contributions from these boundaries, and these do not have to
be at the boundary of $AdS_3$. Thus, it seems that D-brane physics
in the bulk could be manifestly different after the deformation, even
in Euclidean space.
Such physics
could involve for instance D-instanton corrections to correlation
functions, D-branes localized in the bulk, or D3-branes wrapping an
$AdS_2\times S^2$ cycle in $AdS_3\times S^3$.
D-branes in $AdS_3$ have been studied for example
in \refs{\StanciuNX,\FigueroaOFarrillEI,\BachasFR,\MaldacenaKY,
\PetropoulosQU,\GiveonUQ,\LeeXE,\HikidaYI,\RajaramanCR,
\ParnachevGW,\RajaramanEW,\RyangPX}.

Studying this requires us to be able to calculate correlation functions
with \contform\ inserted along the boundary.  In general we do not
know how to treat $k(z)$ and $\Lambda$ near the boundaries of the
worldsheet.  However, in certain circumstances, $\Lambda$
approaches an $x$-dependent constant near the boundary,
and we can calculate the effect of
the deformation explicitly. One such circumstance involves worldsheets which
can be treated semiclassically.  In such a case we
can simply replace $\Lambda$ by the value of \lambdavert\ at the
locus in the target space where the boundary of the worldsheet is
mapped. Another involves D-branes which preserve a diagonal
subgroup of the $SL(2)\times SL(2)\times SU(2)\times SU(2)$
chiral algebra.  In these cases the symmetries determine
the behavior of $\Lambda$ near the worldsheet boundaries.  A third
situation in which we have control is that of D-instantons on
$AdS_3$, which freeze the worldsheet boundaries in all directions.
Here again we can replace the worldsheet fields
$\gamma,\bar\gamma,\phi$ appearing in \lambdavert\ by their
boundary values.  We believe that a similar situation may also
occur for D0-branes on $AdS_3$, at least with regard to emission of
massless closed strings whose worldsheets intersect the D-branes
at a point (up to string scale fluctuations, which may be canceled
by ghosts, since they are just along the longitudinal time direction).

Our goal is to understand the effect of our deformation on the
physics of the D-branes.  This requires studying worldsheets with
boundaries and insertions of \wsdef.
 From the localization of $K$ to a contour integral around each
boundary, we see that in the above cases where $\Lambda$
approaches some constant $\Lambda_i(x,\bar x)$ at the $i$'th
boundary, the expression for $K$ reduces to $\sum_i\Lambda_i
\oint_i {dz\over{2\pi i}} k(z) $, where the sum goes over the
disconnected boundaries of the worldsheet.  The contour integral
produces the charge $q_i$ of the closed string channel state
emitted by the D-brane.
Thus,
the effect of the deformation on a diagram with particular charges
$q_i$ floating through it is to multiply the diagram by
\eqn\insertion{exp({\tilde h \over {Q_1 Q_5}}
\int d^2 x \sum_{i,j} q_i\Lambda_i(x,\bar
x)\tilde q_j\bar\Lambda_j(x,\bar x)).}
Using the fact that the closed string vertices depend on $q_i$
simply through a factor of $e^{iq_i \theta}$ (if we choose $\theta$
to be an angular variable along the isometry generated by $J$) and
on ${\tilde q}_i$ similarly through a factor of $e^{i{\tilde q}_i
{\tilde \theta}}$, one can show that (in the case of constant
$\Lambda$) all string diagrams involving
D-branes sitting at positions $(\theta_k, {\tilde \theta_k})$ are
multiplied by an insertion of the form
\eqn\insertionII{exp(-{\tilde h \over {Q_1 Q_5}}
\int d^2 x \sum_{k,l} \Lambda_k(x,\bar
x) \bar\Lambda_l(x,\bar x)
{\del \over {\del \theta_k}}
{\del \over {\del {\tilde \theta}_l}}
),}
where here the sum goes over the different D-branes in the background
and we are assuming that none of the D-branes lie at fixed points of
the isometries (since the $\theta$'s are ill-defined there).

\fig{Annulus contribution to the force between D-branes
at order $\tilde h g_s^2$.}{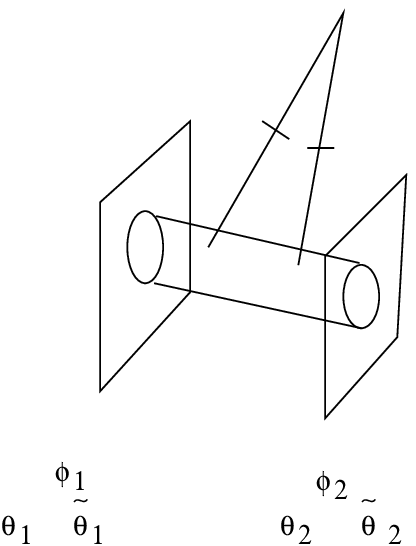}{4 truecm}
\figlabel\annfig

For disk diagrams, with no charged closed string insertions,
the deformation has no effect since no charge can be emitted by the
boundary state (nothing can absorb it, and the contour integral above
can be shrunk to zero size).
Therefore, the leading contribution in all our calculable cases of
D-brane interactions could arise from diagrams at order
$g_s^2\tilde h$.  One such contribution is the annulus with one
insertion of the deformation operator, as depicted
in figure \annfig.  Other contributions at the same order
come from diagrams where the ``double-trace'' wedge connects
two otherwise disconnected annuli.  We can calculate all these
diagrams equivalently using \wsdef\ or \Abdry.\foot{For example,
we can use \Abdry\ to calculate the diagram in figure
\annfig\ as follows.
Let us denote by $Q(y)$ the charged field propagating
in the closed string channel, with charges $q$ and $\tilde q$
under our two U(1)'s.  The amplitude is
\eqn\basicB{
\langle B_1 | \int d^3y q:A_\mu Q\del^\mu Q(y):
\int d^3y^\prime :\tilde q\tilde A_\nu Q\del^\nu Q(y^\prime):
{\tilde h\over{Q_1Q_5}}\int_\del : A\wedge \tilde A :|B_2 \rangle,
}
where $|B_1\rangle $ and $|B_2\rangle $ are boundary states
corresponding to the two D-branes, projected onto the sector with
charges $q$ and $\tilde q$, and where we have pulled
down from the action three interaction terms:  two cubic
couplings between charged fields and the Chern-Simons gauge field,
and the boundary term \Abdry.  All of the fields here can be
contracted with each other (or in the case of two of the
$Q$'s, with the boundary states).  The contraction between
the bulk $A_{\mu}(y)$ and the boundary $A_\del$ gives the
bulk-boundary propagator encoded in the vertex operator
\Jvertex, and similarly for $\tilde A$.  This yields the
diagram in figure \annfig.}

In some cases this
contribution will vanish.  For example, when $\Lambda$ takes
the same value on both boundaries of the annulus, then the sum
over $i$ (or over $j$) in \insertion\ vanishes by charge conservation.
This cancellation occurs
for each closed string charge sector separately. The path integral
involves a sum over all closed strings propagating between the two
boundaries, and in particular a sum over all the possible closed
string charges. Thus, another source of cancellation can arise (for
example) when we deform by the $U(1)$ currents
inside the $SU(2) \times SU(2)$, if the D-branes are not separated
on the $S^3$, since then the sum over positive and negative $q_i$
(and/or $\tilde q_j$) cancels (for a generic position of the D-branes
which is not a fixed point of the isometries).
If we separate the D-branes on the $S^3$ this
cancellation is avoided by having different $q_i$ and $\tilde
q_j$-dependent spherical harmonics appearing in the closed string
wavefunctions emanating from the separated branes.  However, when
these separated D-branes contribute to instanton effects, one
integrates in spacetime over their positions on the $S^3$,
yielding again a cancellation.  In particular, this cancellation
would occur in calculating instanton corrections to the ``moduli
potential'' which we know from the dual CFT must cancel.
We will discuss this further in \S5.2.

We will mostly be interested in studying the effects of
supersymmetry breaking on the bulk D-branes.  In the original
background, there are D-branes which break all the supersymmetry
and therefore have 16 fermionic zero modes on their worldvolume
from the broken supercharges, and there are other branes which
break half the supersymmetry and have eight fermionic zero modes.
We find that all these zero modes can be (and presumably are)
lifted at order $\tilde h
g_s^2$ from the diagram of figure \annfig.  This is a local bulk signal of
supersymmetry breaking, in contrast to the closed
string sector where no such effect arose
semiclassically in the Euclidean case.
We will also study vacuum annulus diagrams,
which indicate the effect of the deformation on forces between
D-branes.  The picture that emerges (at least at leading order
in $\tilde h$) is
that the D-branes do not sit in supermultiplets after the
deformation, but because of the integration over spacetime
collective coordinates, they do not contribute destabilizing
instanton effects.

\subsec{Localized Bulk D-branes}

The $AdS_3\times S^3\times T^4$ background arises as the near-horizon
limit of fundamental strings parallel to NS 5-branes wrapped on $T^4$.
We can imagine putting in additional particle-like
D-branes in this background --
say, in type IIB, D1-branes or D3-branes wrapped around the 1-cycles
and 3-cycles of the $T^4$. Before we took the near-horizon limit, these
D-branes were attracted to the F1-NS5 system,
and they could form a bound state whose energy was the square root of
the sum of the energies squared of the separate systems (which is the
BPS bound; the bound state is supersymmetric). If the F1-NS5 system
is wrapped on a circle, the additional D-branes have a finite contribution
to its energy, while if it is on a line they do not contribute to it. Thus,
after taking the near-horizon limit, we find \LarsenUK\ that in
Poincar\' e coordinates there is no lower bound on the mass of D-branes,
but there is such a bound in global coordinates. This bound, which is
proportional to the number of D-branes squared, appears even though the
D-branes break all the supersymmetry; it is related to the original
supersymmetries of the F1-NS5 system which are non-linearly realized.
In any case, at weak coupling it is easy to see that
such D-branes in $AdS_3\times S^3
\times T^4$ have a mass which is much larger than the lower bound (this
is fortunate since, for small D-brane number when we can ignore
back-reaction, the mass grows linearly with the number of D-branes),
they break supersymmetry completely, and one expects to have generic
forces between them in the bulk (which at large distances arise from
the exchange of massless particles). Moreover, these branes are not static
in the bulk of $AdS_3$, but rather follow the geodesics
for massive particles. In our coordinate system this means they are
attracted towards smaller values of $\phi$. This motion is insignificant
at time scales much smaller than $L_{AdS}$, and in our discussion
we will assume we are dealing with such time scales and we will ignore it.
In addition to such branes which are
D0-branes on $AdS_3$, we could also consider D-instantons on $AdS_3$,
such as the type IIB D-instanton or Euclidean D-branes wrapped on
cycles of the $T^4$. These also completely break the supersymmetry.

Let us consider the annulus contribution of figure \annfig\ in the case
that the two boundaries are
localized on $AdS_3$.  We place the D-branes, or the boundaries
of the annulus, at
positions $y_i=\{ \gamma_i,\bar\gamma_i,\phi_i \}$ on $AdS_3$
and $\theta_i,\tilde\theta_i$ on the two circles
on the $S^3$
corresponding to $J$ and $\tilde J$, where $i=1,2$ labels
the two branes.
We will use the semiclassical equation
for $\Lambda$,
\eqn\semilam{
\Lambda_i=\Lambda_{i, semiclassical}=-{{(\bar\gamma_i-\bar x)e^{2\phi_i}}
\over{1+|\gamma_i-x|^2e^{2\phi_i}}}.
}
For D-instantons, the boundary of the worldsheet
cannot fluctuate since there are Dirichlet conditions in
all directions.  In this case
we also find that the semiclassical expression \semilam\
agrees with the expression for $\Lambda$ in \GiveonUQ, where it
was found for a particular boundary condition
that near the boundary an operator $\Phi_1$, which is related
to the operator $\Lambda$ by $\del_{\bar x} \Lambda = \pi \Phi_1$, goes
to a constant times $1/(1+|x|^2)^2$ as we approach the boundary\foot{In
fact, in \GiveonUQ\ various different possible boundary conditions were
discussed, which give somewhat different behaviors of $\Phi_1$ near the
boundary. From an analysis of the symmetries of the problem it seems
clear that the form of $\Phi_1$ above must be the one corresponding to
D-instantons, though this is not what is claimed in \GiveonUQ.}. This
leads to $\Lambda \to {\bar x}/(1+|x|^2)$, which exactly agrees with
our expression above for an instanton positioned at $\gamma={\bar
\gamma}=\phi=0$, which is the instanton corresponding to the boundary
conditions discussed in \GiveonUQ\ (other instantons can be generated
from this by $SL(2)$ transformations).
In the case of D0-branes,  the boundary of
the worldsheet can fluctuate in at most one (timelike) direction.
We expect this longitudinal fluctuation to be cancelled
by ghosts (and in the case of heavy winding mode exchange,
to be suppressed regardless).

For simplicity let us take the two boundaries at
$\gamma_i=\bar\gamma_i=0$ and place the D-branes at
points on $S^3$ which are not fixed points of the isometries
corresponding to $J$ and $\tilde J$.
Note that by charge conservation along
the diagram, $q_1=-q_2=q, {\tilde q}_1 = -{\tilde q}_2=\tq$.
Working at first order in $\tilde
h$, plugging \semilam\ into \insertion, we obtain a contribution
of the form
\eqn\locbasic{ {\cal A}_{q,\tilde q}={\tilde h \over {Q_1Q_5}}
\int
d^2x q\tilde q \biggl| {{\bar x
e^{2\phi_1}}\over{1+|x|^2e^{2\phi_1}}} -{{\bar x
e^{2\phi_2}}\over{1+|x|^2e^{2\phi_2}}}\biggr|^2
G^{(0)}_{q,\tilde q}(\theta_i,y_i) }
to the
annulus amplitude arising from closed strings exchanged with
particular $U(1)\times U(1)$ charges $(q, \tilde q)$, where
$G^{(0)}$ gives the annulus contribution without our ``double-trace''
insertion. The angular dependence of this contribution is of the form
\eqn\phases{
e^{i(q_1\theta_1+\tilde q_1\tilde \theta_1)}
e^{i(q_2\theta_2+\tilde q_2\tilde\theta_2)}=
e^{iq(\theta_1-\theta_2)}e^{i\tilde
q(\tilde\theta_1-\tilde\theta_2)}, }
due to the
wavefunctions of the closed strings at the two ends of the
annulus.  These contributions \phases\ explicitly break the
symmetry which would otherwise exist between positive and negative
values of $(q,\tilde q)$.  Note that in the absence of these
contributions (for instance, if $\theta_1=\theta_2$ or $\tth_1 =
\tth_2$), the contributions from positive and negative
$q,\tilde q$ in \locbasic\ would
cancel when we sum over the different charge sectors.

The $x$ integral in \locbasic\ can be performed, yielding
the result
\eqn\doneint{
{\cal A}_{q,\tilde q}={\tilde h G^{(0)}_{q,\tq}
\over {Q_1 Q_5}} q\tilde q
\bigl(-2+2(\phi_1-\phi_2) \coth(\phi_1-\phi_2)\bigr).
}
For fixed nonzero separations $\theta_{12},\tilde\theta_{12}$,
this contribution survives the sum over $q,\tilde q$.
This result constitutes a contribution to the force
between D-branes (or in the D-instanton case, to the
instanton action) which is present in the bulk of AdS.
Because of the power of $1/L_{AdS}$ implicit in the
$\phi_{12}$ contributions, with our current scalings
this force disappears in the flat space limit
$L_{AdS}\to \infty$, which is the same limit
in which the $AdS_3$-induced tadpoles for the positions of the
D-branes disappear.
It therefore agrees nicely with the
type of contribution expected from the CFT side.
In the next section we will discuss another scaling
for $\tilde h$ in which these contributions in fact
survive in the flat space limit.

We can similarly calculate contributions from the other
diagrams at order $\tilde h g_s^2$, involving two annuli connected by the
deformation.  For the D-instanton
case, this leads to a similar contribution to \doneint; now we have
four charges characterizing the diagram, $(q,\tq)$ flowing through
one annulus and $(q',{\tq}')$ flowing through the other, and the result
is
\eqn\doneinttwo{
{\cal A}_{q,\tilde q,q',\tq'}={\tilde h G^{(0)}_{q,\tq} G^{(0)}_{q',\tq'}
\over {Q_1 Q_5}} (q \tq' + q' \tq)
\bigl(-2+2(\phi_1-\phi_2) \coth(\phi_1-\phi_2)\bigr).
}
These contributions thus give different
$\phi,\gamma,\bar\gamma,\theta,\tilde\theta$-dependence than
the one we calculated above.

\fig{Annulus contribution to the mass of D-brane worldvolume
fermions at order $\tilde h g_s^2$.}{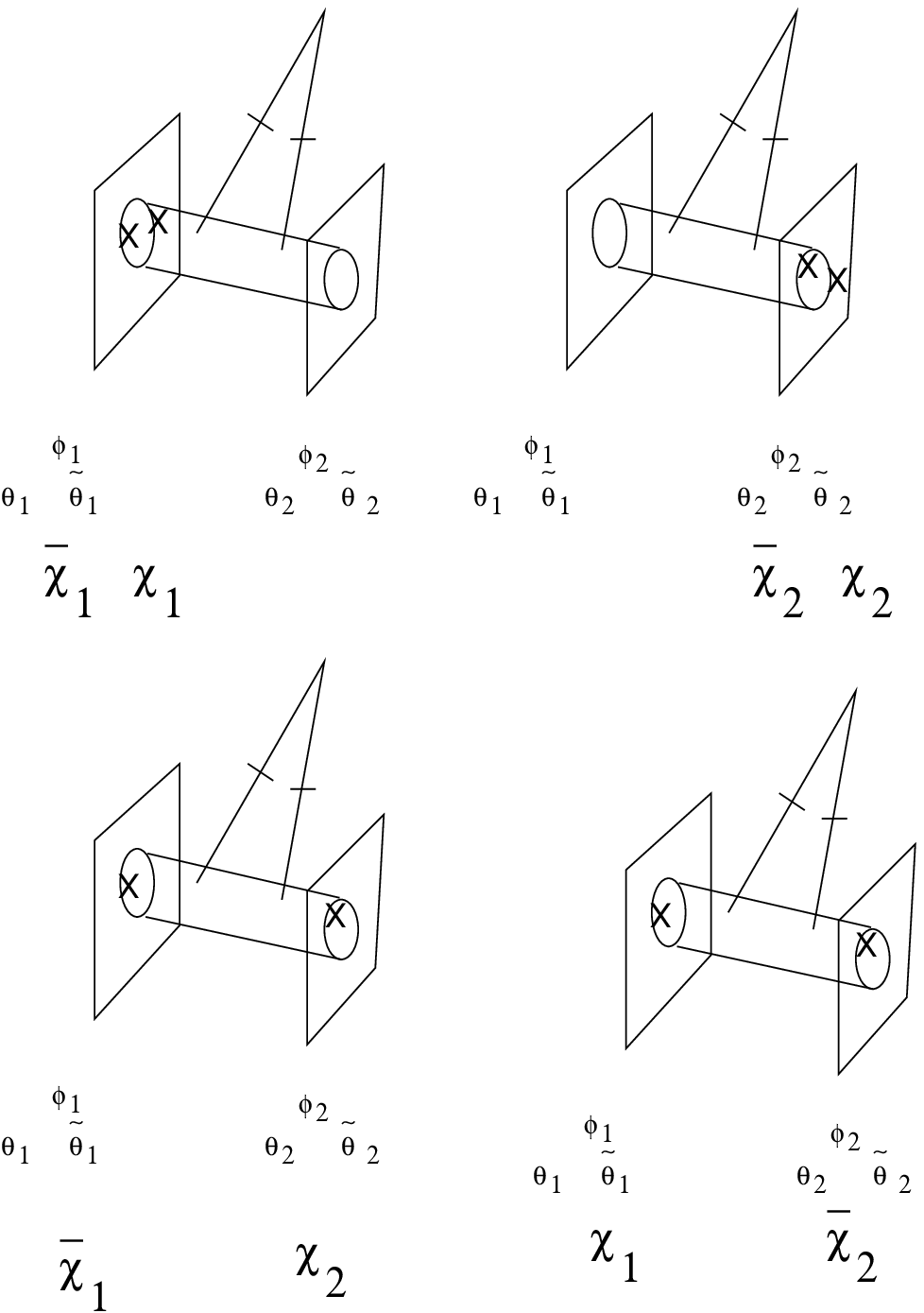}{10 truecm}
\figlabel\zmfig

A very similar calculation predicts the lifting of the worldvolume
fermion zero modes (Goldstinos) of the pair of D-branes.  Before
our ``double-trace'' deformation is turned on, space-time
supersymmetry (in the absence of D-branes) is unbroken and the
system of D-branes sits in a long multiplet and has 16 fermionic
zero modes which are responsible for creating its superpartners.
Let us denote the fermion zero modes on the $i$'th brane
$\chi_i,\bar\chi_i$.  Before the ``double-trace'' deformation, the
quadratic terms for these fields on the worldvolume of the pair of
branes are of the form
\eqn\premasses{
(\bar\chi_1-\bar\chi_2)(\chi_1-\chi_2),}
so that the overall
combinations $\chi_1+\chi_2$ are massless\foot{We are being schematic here,
and ignoring the various indices of the fermions and the dependence of
the massless combinations on the positions of the D-branes.}. The issue is then
whether the contributions in figure \zmfig, which are the leading
corrections to the fermion masses, produce the same
combination of quadratic terms, preserving the masslessness of
$\chi_1+\chi_2$, or not.  It is easy to convince oneself that
there is no reason why the order $\tilde h$ amplitude should
produce a result proportional to the combination \premasses.  This
is because the charges propagating in the closed string channel of
the diagram are different for diagrams with one fermion on each
boundary (which contribute masses
$\bar\chi_1\chi_2,\bar\chi_2\chi_1$) relative to those with two
fermions on a single boundary (which contribute masses
$\bar\chi_1\chi_1,\bar\chi_2\chi_2$). The first two diagrams in
figure \zmfig, with two fermions inserted at a single boundary of
the annulus, yield a contribution of the form \locbasic\ with a
sum over integer $q,\tilde q$. The last two, with fermions on
different boundaries, have fermionic closed strings propagating in the
diagram, so (when the deformation involves the $U(1)_R$ currents) they
involve a sum over half-integer $q,\tq$.
Therefore, we do not expect the combination
\premasses\ where the two types of diagrams are weighted the same
to persist at order $\tilde h$, and we expect all fermion zero
modes to be lifted.

Thus, we have determined a bulk supersymmetry breaking
effect of our NLST deformation in this system, at the
level of forces between D-branes in the theory and their worldvolume
action.

\subsec{Nonperturbative Nonrenormalization in Nonsupersymmetric
Non-local String Theory}

As we explained above, an interesting feature of our
deformation is that it breaks supersymmetry without introducing
destabilizing tadpoles for moduli.  From the field theory
side, this is an exact statement.  It is interesting therefore
to explore how this phenomenon arises on the gravity side,
given that we have just manifested bulk SUSY breaking effects
in the D-brane sector.

In order to do this, there is a step remaining in the calculation.
D-branes contribute to the ``moduli potential'' via virtual loops
and instanton effects, which require a second quantized spacetime
description.
In such a calculation, \doneint\ can represent
a correction to the instanton action.  The effect of the instanton
on physical quantities in spacetime is obtained by a spacetime
path integral including integrals over all the fermionic
and bosonic zero modes.  The fermionic zero modes, which before
the deformation caused the amplitude to vanish,
are now lifted.  However, the bosonic zero
modes, including the positions $\theta_i,\tilde\theta_i$, remain.
Although we get a contribution for each value of
$\theta_i,\tilde\theta_i$ as discussed above, the integral
over these zero modes of \doneint\ cancels due to the phases \phases.
Similarly, the diagrams we computed in \doneinttwo\ cancel after
integration over the positions unless $q=-q'$ and $\tq=-\tq'$, and
the remaining amplitudes cancel when we sum over the possible values
of $q$ because of a cancellation between positive and negative $q$'s.

At this order, this provides a satisfying resolution to the
problem of how the gravity side manages to avoid generating
a ``moduli potential'' despite the supersymmetry
breaking introduced by the deformation (and the absence of
fermion zero modes).  The D-branes
experience non-local SUSY breaking forces in the bulk, but
these effects cancel in computing their virtual and
instantonic contributions to other physical observables
via a cancellation in the integration over
bosonic zero modes $\theta,\tilde\theta$.

We can apply this result from the D-brane
sector to get more intuition, at least heuristically, for
the cancellation of the ``moduli potential'' in the closed
string sector discussed in \S3.2.  A diagram with charged
closed strings running in loops would naively seem to
contribute to the ``moduli potential'' once the deformation
which splits their masses according to \YdimII\ is turned
on.  However, at the worldsheet level we have seen that
semiclassically (in Euclidean space)
the vertex operators $K,\tilde K$ localize
on boundaries and charged vertex operator insertions,
introducing factors of the form \insertion\ into
the contributions of individual worldsheets with
charges $q,\tilde q$ propagating from boundaries
or vertex operator insertions.  The moduli are uncharged,
so from the worldsheet point of view it is clear that
the closed string ``moduli potential'' still cancels also after the
deformation.

\fig{Degenerating Riemann surface contributing cancelling
contributions to the ``moduli potential''.}{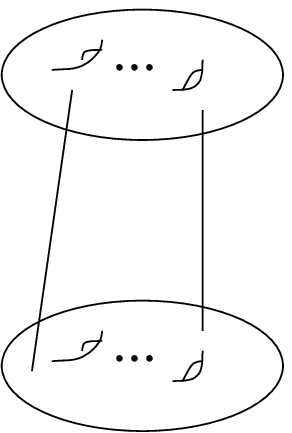}{3 truecm}
\figlabel\degfig

However, we can dissect the closed string
diagrams in a way that
provides a little more intuition for how the naive spacetime
intuition fails in this non-local theory.  Consider
a Riemann surface $\Sigma$ which has degenerated into separate
Riemann surfaces $\Sigma_i$ connected by a set of thin tubes, as in
figure \degfig.
The ends of the tubes can be approximated by local operator
insertions $T_{ij}(z,\bar z)$ on the $\Sigma_i$.  The $K$ and $\tilde K$
insertions on each $\Sigma_i$ then localize on the
insertions $T_{ij}$, and for diagrams in which the closed
strings propagating in the long thin tubes are charged,
one gets a contribution from this.

Semiclassically, at order $\tilde h$, one therefore gets an
insertion of the form \insertion\ where the $\Lambda_i,
\bar\Lambda_j$'s are the values of $\Lambda,\bar\Lambda$ at the
positions of the ends of the long thin tubes.  Generically, a
semiclassical analysis will not be valid, but in some
circumstances (such as when the strings propagating in the
$\Sigma_i$ are very heavy from say winding or momentum along the
$T^4$) it will be.  In any case it gives a useful heuristic
picture of how cancellations might occur in spacetime similarly to
the case of D-branes. Namely, the contribution to the ``moduli
potential'' again involves integrating over the positions
$\theta,\tilde\theta$ of the insertion points of the tubes, giving
a cancellation at order $\tilde h$.

It is not obvious from the point of view described in this section
what happens to the D-brane corrections to the ``moduli
potential'' on the gravity side at higher orders in $\tilde h$ or in $g_s$.
The field theory side
again predicts no contributions to the ``moduli potential''.
There are several diagrams at order $\tilde h^2$ which
must therefore cancel if the duality is correct.  These cancellations
may be nontrivial, analogous to two and higher loop cancellations
of protected quantities in supersymmetric theories which
do not follow from any simple counting of Bose-Fermi
degeneracies.  In our case,
the only symmetry principle we have so far
identified to enforce the cancellation
is the duality (namely, the exact marginality of the deformation on
the field theory side), and it would be nice to obtain a more direct
argument applicable for arbitrary $\tilde h$ on the gravity side.

\newsec{The Flat Space Limit}

It is interesting to contemplate NLST's arising in backgrounds other
than AdS.  One way to try to construct such backgrounds is to consider
the flat space limit of the AdS realizations we have so far. It seems
that we should not expect such a limit to make sense, since our
deformation is maximally non-local on the $S^3$, and
induces correlations at distances of the order of the AdS
scale that go to infinity in the flat limit, leading to failure of the
standard conditions for unitarity.  This is related to the fact that in
taking the flat limit one focuses on one small region of the $S^3$, and
the other regions which are correlated with it in the original
theory go off to infinity.
In this section we will show that there is a scaling of $\tilde h$ which gives
finite contributions when one takes the $L_{AdS}/l_s\to \infty$ flat
space limit of the results derived
in the previous section, and also gives a finite
non-local deformation of the worldsheet action in the same limit. However,
we have not been able to find sensible vertex operators in the resulting
theory, so it is not clear if the flat space limit defines a sensible (unitary)
NLST or not.

\subsec{Definition of the Flat Space Limit}

The flat space limit of $AdS_3$ backgrounds with NS-NS charges
involves
taking $Q_1$ and $Q_5$ to infinity with a fixed ratio $Q_5/Q_1=g_6^2$.
Since the $AdS_3$ string metric goes
as $ds^2 = Q_5 (d\phi^2 + e^{2\phi} d\gamma
d{\bar \gamma})$, the relation between $\phi$ and a flat space
coordinate $X_{\phi}$ is of the form $\phi \simeq X_{\phi}/\sqrt{Q_5}$.
Thus, if we wish to keep $X_{\phi}$ constant (which is the simplest
possibility) we need to take
$\phi \to 0$ when we take the flat space limit.
Similarly, when we expand around some particular generic point on
$AdS_3\times S^3$, the angular coordinates on the $S^3$ are related to
flat space coordinates by $\theta \simeq X_{\theta}/\sqrt{Q_5}$,
${\tilde \theta} \simeq X_{\tilde \theta}/\sqrt{Q_5}$, and the
charges $q,{\tilde q}$ become momenta $p,{\tilde p}$ in the
$X_{\theta}, X_{\tilde \theta}$ directions, where $p=q/\sqrt{Q_5}$,
$\tilde p = {\tilde q} / \sqrt{Q_5}$.

Consider for example \doneint. In the flat space limit
this result reduces to
\eqn\loclim{ {\tilde h \over {Q_1 Q_5}} q_1\tilde q_1 G_0
{4\over 3}(\phi_1-\phi_2)^2 = {\tilde h \over {Q_1 Q_5}}
p_1 {\tilde p}_1 G_0 {4\over 3} (X_{\phi_1}-X_{\phi_2})^2. }
Therefore if
$\tilde h$ is constant, independent of $Q_5$, then this effect
disappears in the limit (we are assuming that the amplitude $G_0$
before the deformation has a finite flat-space limit). We want
the effect to actually depend in the
flat space limit only on $g_6^2 = Q_5/Q_1$. Thus, we need to take $\tilde h
\to \infty$ as
\eqn\flath{
\tilde h=\tilde h_0 Q_5^2, }
where $\tilde h_0$ is constant, and then we
get a finite surviving contribution in this limit.

Let us denote the position of one brane by $X$
and the other by $Y$. Then, because of the factors \phases\ and \loclim,
the order $\tilde h$ contribution to the
annulus diagram for a particular closed string $s$ exchanged in the flat
space limit is proportional to
\eqn\flatans{
\partial_{X_{\theta}}\partial_{X_{\tilde \theta}}D_s(X-Y),
}
where $D_s(X-Y)$ is the contribution of this mode to the exchange
force and we only wrote down the dependence on $X_{\theta},X_{\tilde
\theta}$
(for a graviton exchange diagram $D_s$ is the position-space propagator
between the D-branes).  In the flat space limit, the sum
over charges $q,\tilde q$ turns into a continuous integral over
momenta $p, \tilde p$ in the $X_{\theta}, X_{\tilde \theta}$
directions.  This washes
out the supersymmetry breaking effects, which arise from the
distinction between sums over $q,\tilde q\in \IZ$ and
sums over $q,\tilde q\in \IZ+1/2$.  So the force between flat space
BPS branes will cancel when all the contributions are added in (since
the added contributions will still be supersymmetric),
but for branes and anti-branes the force discussed above will persist in the
limit.

It is instructive to spell out more explicitly the form
of the vertex $K(x)$ appearing in the deformation \wsdef\
in the flat space limit.  Taking the limit as in \flath,
with $\tilde h$ scaling as $Q_5^2$, the deformation is
\eqn\defagain{
\delta S_{ws} = {\tilde h}_0 g_6^2 \int d^2x K(x)\tilde K(\bar x).
}
Taking the limit as above, one finds (from \Jvertex\ and \lambdavert)
\eqn\vertlim{
K(x)\to {1\over \pi} \int d^2 z \del_z X_{\theta}
\biggl[{{-2\bar x}\over{(1+|x|^2)^2}}\del_{\bar z}X_{\phi}
+{1\over{(1+|x|^2)^2}}\del_{\bar z} X_{\bar\gamma}
-{{\bar x^2}\over{(1+|x|^2)^2}}\del_{\bar z} X_{\gamma} \biggr],
}
where $X_{\phi}=\sqrt{Q_5}\phi, X_{\gamma}=\sqrt{Q_5}\gamma,
X_{\bar\gamma}=\sqrt{Q_5}\bar\gamma$ are the flat
space coordinates descending from the $AdS_3$ coordinates
as discussed above, and similarly for $\tilde K({\bar x})$.
This linear combination of
$\del_{\bar z} X^{\mu}$ descends from a longitudinal (formally
pure gauge) vector potential in $AdS_3$, and does not
have fermionic pieces as a result\foot{We thank D. Kutasov
for a discussion on this point.}.  In the flat space limit, $K(x)$
is an integrated
physical vertex operator for a tensor field in spacetime at
zero momentum.

Plugging \vertlim\ into \defagain\ and performing the integral over
$x$, we obtain
\eqn\defflat{\eqalign{
\delta S_{ws} \propto
\tilde h_0 g_6^2 &\ \int d^2z_1\int d^2z_2
\biggl[2 \bigl(\del_{z_1}X_\theta\del_{\bar z_1}X_\phi
\bigr)\bigl(\del_{\bar z_2}X_{\tilde\theta}\del_{z_2}X_\phi\bigr)\cr
&+ \bigl(\del_{z_1}X_\theta\del_{\bar z_1}X_{\bar\gamma}
\bigr)\bigl(\del_{\bar z_2}X_{\tilde\theta}\del_{z_2}X_\gamma\bigr)
+ \bigl(\del_{z_1}X_\theta\del_{\bar z_1}X_{\gamma}
\bigr)\bigl(\del_{\bar z_2}X_{\tilde\theta}\del_{z_2}X_{\bar\gamma}
\bigr)\biggr]. \cr
}}
Note that the coefficients in front of the three terms are exactly
those which give an $SO(3)$ rotational invariance in the $X_{\phi},
X_{\gamma},X_{\bar \gamma}$ directions, as expected in the flat space
limit (in the Lorentzian case this will become $SO(1,2)$).

Thus we obtain a deformation of the general form \genwsdef\ which
persists in flat space.  The deformation we have discovered
is very simple: it consists of
a sum of bilocal products of linear combinations of zero-momentum
off-diagonal graviton and NS B-field vertex operators.
Since they are total derivatives, these vertex operators
localize to the boundaries
of the worldsheet or to other operator insertions.
The NS B-field decouples from
closed strings, and the off-diagonal metric couples
to modes with momentum along the $X_\theta$ and
$X_{\tilde\theta}$ directions.

\subsec{Observables in the Flat Space Limit ?}

We would like to study whether the theory we
obtain in this limit is sensible. To do so it is important to
formulate and study the behavior of physical observables in this
theory.  Because of the relative simplicity of the theory \defflat, we
can investigate this question rather explicitly.
We will consider two types of candidate
observables, using two techniques for analyzing the deformed
theory. The first arises by considering familiar flat space vertex
operators inserted into the path integral with the bilocal
contribution to the action
\defflat. The second, described in \S6.3,
arises by considering a different but equivalent
presentation of the theory, in terms of a Lagrange multiplier
which renders the action Gaussian, and considering a particular
set of non-local insertions in the path integral which are natural
in this formalism.  In both cases, because of the non-locality of
the underlying theory, we will find in the end no separately
renormalizable constituents in a given amplitude; instead we will
be left with a rather unpredictive situation in which each
amplitude must be independently renormalized. This is presumably related
to the problems one expects with unitarity when taking a limit which
keeps only a region much smaller than the non-locality scale.

In the first approach we calculate correlation functions of vertex
operators in the
flat space limit by inserting powers of \defflat\ to obtain the effect of
our deformation, and we find that this leads to divergences. Consider for
example a correlator of $n$ vertex operators $V_{p_j}\sim
e^{ip_j\cdot X}$.
Let us compute the order $\tilde h_0$ correction to
this correlator coming from the first term of our deformation
\defflat. It is given by
\eqn\flatcorr{\eqalign{\tilde h_0 \int \prod_{k=1}^n d^2w_k d^2z_1
d^2z_2 \vev{\prod_{j=1}^n & e^{ip_j\cdot X(w_j,\bar w_j)} \del
X_\theta\bar\del X_\phi(z_1,\bar z_1) \del X_\phi\bar\del
X_{\tilde\theta}(z_2,\bar z_2)} \sim\cr \sim \tilde h_0
\int\prod_{k=1}^n d^2w_k d^2z_1 & d^2z_2 \prod_{i,j=1}^n
|w_{ij}|^{p_i\cdot p_j/2}
\cdot \cr \cdot
& \biggl[\sum_{i=1}^n{{p_i^\theta}\over{z_1-w_i}}\biggr]
\biggl[\sum_{i=1}^n{{p_i^\phi}\over{\bar z_1-\bar w_i}}\biggr]
\biggl[\sum_{i=1}^n{{p_i^\phi}\over{z_2-w_i}}\biggr]
\biggl[\sum_{i=1}^n{{p_i^{\tilde\theta}}\over{\bar z_2-\bar
w_i}}\biggr].\cr }}
The last four factors come from contractions of the zero-momentum
vertex operators in the deformation with those of the $n$ vertex
operators whose correlation function we are calculating.  The
integrals over $z_1$ and $z_2$ diverge when a zero-momentum vertex
operator hits an $e^{ipX}$ on the worldsheet.  In ordinary flat
space string theory, this divergence is a standard pole in the
S-matrix arising from the fact that when a zero-momentum particle
combines with a momentum $p$ particle to produce a momentum $p$
particle, the latter is still on-shell and gives a pole (this
can be seen explicitly by continuing the zero momentum vertex
operators to nonzero momentum $q$ and expanding in small $q$).  We
would like to understand the meaning of this divergence in our
application, where this correlator describes the shift of the
correlation function of vertex operators $V_{p_i}$ under the NLST
deformation.

Let us first regularize this divergence. If we put a
short-distance cutoff on the worldsheet analogous to \wsa\ in the
AdS case, namely letting other operators approach only to a
distance $a_j$ from $V_j$, we find that we need to redefine :
\eqn\birescale{\biggl[ \prod_{j}\int d^2w_jV_j(w_j) \biggr] \to
\biggl[ \prod_j\int
d^2w_jV_j(w_j) \biggr] \biggl(1-\sum_{l,k}\tilde h_0 g_6^2 p_l^\theta
p_l^\phi log|\tilde a_l|^2 p_k^{\tilde\theta} p_k^\phi
log|\tilde a_k|^2 \biggr),}
where $\tilde a$ is proportional to $a$ and absorbs some
subleading contributions.
This shift cancels the divergence above at leading order in
$\tilde h_0$. Note that the shift we need for the product of vertex
operators is not equal to the product of the shifts we need for each
vertex operator separately.
This would not occur in a local worldsheet string theory.
However, since in a NLST the worldsheet
Lagrangian is non-local, it may be necessary
to consider as observables the full
set of multilocal excitations of the theory, since attempting to
consider only local vertex operators would generically fail under quantum
corrections.

Unfortunately,
this prescription appears to render the theory unpredictive as far as these
observables go, since one must renormalize separately each
physical process rather than obtaining predictions for physical
processes arising from a finite number of renormalizations of
constituent fields and couplings.
It is therefore unclear whether the theory is
renormalizable in the appropriate sense, because each combination
of vertex operators is a new multilocal operator in the theory and
one therefore has to input an infinite amount of information to
define the set of observables. Because of this issue, our results on the flat
space limit are inconclusive (though we think intriguing) and
we hope to improve our understanding of the proper physical
constraints on this sort of theory in general backgrounds in
future work.

We started with a theory in which the non-locality scale
is of the order of the AdS curvature radius $L_{AdS}$,
and this goes to infinity in the flat space limit.
It would be very interesting to
figure out what (if any) are the appropriate observables in such a non-local
theory, that can give meaningful physical amplitudes.
Of course it is worth emphasizing that with $\tilde h$ scaling
independently of $Q_5$, we would obtain conventional flat space
string theory in the limit.
In usual flat space string theory we can define observables
by S-matrix elements describing particles which are much farther from
each other than the characteristic non-locality scale. These
observables give well-defined correlation functions. In the flat space
NLST's we constructed in this section we have seen that this fails, so
some other types of observables are needed in order to get physical
predictions. In the AdS case the consistency of our NLST constructions
was guaranteed by the consistency of the dual conformal field theory,
but it is not clear what are the consistency conditions for flat space
NLST's. Thus, in the absence of predictions for physical observables
we cannot say if the theories we constructed in this section are
consistent (e.g. if they are unitary) or not.

Although they may render the question of the existence of a useful
flat space limit questionable, the above divergences do teach us
something significant about the $AdS_3$ model that is our main
focus in this paper.  In \S3, we saw that the vertex operators
involved in Euclidean closed string amplitudes localize to the
boundary of $AdS$.  The nontrivial (divergent) answers we find in
the closed string sector after taking the flat space limit here indicate
that there was bulk physics in the closed string sector in AdS.
In particular, as we have seen in some detail, the flat space
limit does not leave us with a consistent S-matrix, which should have been
the case if all of the effects of the deformation were
at the boundary. This provides evidence that the effects
of the deformation, and in particular the non-locality of the
theory, permeate the bulk of AdS space, as expected from the
marginality of the deformation, despite the fact that we can write
the deformation as a boundary term \Abdry.

Note from \flatcorr\ that we see the non-local effects in
the flat space limit only for correlators including vertex
operators with nonzero momentum along what used
to be the $S^3$ directions:  $p^\theta\ne 0\ne p^{\tilde\theta}$.
This is consistent with our expectations from the form
of the deformation \Abdry\ that the $6d$ theory is
non-local even though the effect on the
$3d$ action is a local boundary term.

\subsec{Another Set of Non-local Operators in NLST}

Despite the above complications, one might hope that the physics
simplifies in terms of some other natural subset of observables.  There
is a way of presenting the theory \defflat\ (and more generally
the theories \genwsdef) which simplifies the analysis
considerably, and which suggests another set of multilocal
operators in the theory.

Consider the worldsheet path integral for the theory \defflat,
written as a Gaussian using Lagrange multipliers $\lambda$ (and
ignoring the fermionic fields which play no role) :
\eqn\wslambda{
Z_{NLST}= \int d\lambda \int [DX] e^{-\int d^2z \del X^\mu
G_{\mu\nu}(\lambda)\bar\del X^\nu} e^{-{1\over{2\tilde
H}}\lambda_{\theta\phi}\lambda_{\tilde\theta\phi} -{1\over{\tilde
H}}(\lambda_{\theta\bar\gamma}\lambda_{\tilde\theta\gamma}
+\lambda_{\theta\gamma}\lambda_{\tilde\theta\bar\gamma})}, }
where
$\tilde H \propto \tilde h_0 g_6^2$ and where
\eqn\metric{
G_{\mu\nu}(\lambda)dx^\mu dx^\nu=\eta_{\mu\nu}dx^\mu dx^\nu+
\lambda_{\theta\phi}dx^\theta dx^\phi + \dots, }
where $\dots$ are
other similar terms involving the other $\lambda$'s.
By integrating over $\lambda$ one can see that equation \wslambda\
gives a description of the theory equivalent to the
bilocal description of \defflat, but now the worldsheet path
integral is Gaussian.  This is similar to what arises in
wormhole physics \refs{\HawkingMZ,\GiddingsCX,\ColemanTJ} and
it would be interesting to explore further the conceptual interpretation
of this mathematical trick.

This method seems potentially useful, particularly in our flat space
limit where it renders the partition function Gaussian. As discussed
in section 5 of \AharonyPA\ one can also employ this method in the
$AdS/CFT$ case, but generically in $AdS/CFT$ it is not trivial to
implement either on the gravity side or on the field theory side of
the correspondence.  Naively it simplifies the analysis to one
involving only ``single-trace'' deformations, but in fact this is
complicated on both sides of the duality.  On the field theory side,
the ``single-trace'' operators in question are relevant operators, and
one would be attempting to integrate over the corresponding
scale-dependent couplings. This involves a sum over field theories
with different parameters, whose physical interpretation is
unclear. On the gravity side, these relevant perturbations deform the
geometry dramatically.  In terms of the worldsheet string theory, the
BRST-invariance condition for vertex operators changes as a function
of $\lambda$, an issue we will also encounter in our flat space
analysis here.

Considering just the closed string sector, which feels only
the symmetric part of $G_{\mu\nu}$, we can change variables to
$Y^\mu(z,\bar z)\equiv E^\mu_\nu(\lambda) X^\nu (z,
\bar z)$, where the matrix $E$ is defined by $E_{\mu}^{\rho}(\lambda)
\eta_{\rho \sigma}
E^{\sigma}_{\nu}(\lambda)=
G_{\mu \nu}(\lambda)$.  Then, the path integral becomes
\eqn\newPI{ \int d\lambda \int[DY]\prod_z [\det E(\lambda)]^{-1}
e^{-{1\over{2\tilde
H}}\lambda_{\theta\phi}\lambda_{\tilde\theta\phi} -{1\over{\tilde
H}}(\lambda_{\theta\bar\gamma}\lambda_{\tilde\theta\gamma}
+\lambda_{\theta\gamma}\lambda_{\tilde\theta\bar\gamma})} e^{-\int
d^2 z \del Y^\mu\eta_{\mu\nu}\bar\del Y^\nu}. }
Here the $\lambda$ dependence
is only in the Jacobian (and in the Gaussian),
which is in this flat space situation
independent of the embedding coordinates $Y(z,\bar z)$.

Now let us consider observables (correlation functions of vertex
operators).  A new set of multilocal operators in the $X$ variables
are the simple operators
\eqn\propvert{V_k[Y]\equiv e^{ik\cdot Y}.}
In terms of
$X$, these vertex operators vary with $\lambda$ so as to preserve
conformal invariance in the path integral for arbitrary $\lambda$.
We can insert these into the integrand of
\newPI, and divide by the vacuum path integral \newPI\ to
normalize.  This reproduces the correlators of momentum modes for
ordinary flat space string theory.

These momentum modes \propvert\ of $Y$ are highly
non-local when expressed in terms of $X$ (in the original formulation
of the theory without $\lambda$).
In general, when we map a product of the $V_k[Y]$ operators to the
$X$ variables it will not map into the product of the maps of
these operators.  So in terms of $X$ there is still no
S-matrix with amplitudes determined by
renormalizations of operators describing separated excitations.
However it is true here that there is a set
of multilocal amplitudes (insertions of products of operators
\propvert ) which are naturally determined by the standard
renormalizations of \propvert\ in the $Y$ variables,
and which produce results isomorphic to the flat space S-matrix.

A similar analysis using the \wslambda\ prescription can be
performed in $AdS_3\times S^3\times T^4$, with similar results
arising at leading order in $\tilde h$.  The observables analogous
to \propvert\ there reproduce the standard AdS correlators in the
original undeformed supersymmetric background.  Again they are
non-local and non-locally renormalized in terms of the physical
variables $\phi,\gamma,\bar\gamma,\theta,\tilde\theta$. The meaning of
these observables is unclear, since the physics of the CFT does seem to
depend on $\tilde h$. It is
tempting to speculate that these objects could realize a hidden non-local
supersymmetry in the system which explains the vanishing of the
``moduli potential'', while as we have seen
the physics in terms of the ordinary
variables exhibits broken supersymmetry.

In general, it is important to clarify what are the conditions for
physically consistent
NLST models, both for conceptual interest and with regard to
the potential for applications.  In particular, it would be
very interesting to develop more realistic models that have
the exact stability after supersymmetry breaking that we have
found in the $AdS_3$ backgrounds studied in this paper.

\vskip 1cm

\centerline{\bf Acknowledgements}

We would like to thank D. Kutasov and J. Maldacena for
many helpful discussions.  We would also
like to thank T. Banks, D. Freedman, A. Giveon, S. Kachru, F.
Larsen, A. Lawrence, E. Martinec, H. Ooguri, H. Robins,
A. Schwimmer, S. Shenker, and L. Susskind for useful
discussions.  E.S. would like to thank the
hospitality of the Weizmann Institute during the initial stages
of this work, O.A. and M.B. would like to thank the Benasque Center for
Science, Stanford University
and SLAC for hospitality, and all authors would like to thank the Amsterdam
summer workshop for hospitality during some of its progress,
and the Israel-U.S. Binational Science Foundation for
support. O.A. would also like to thank the Aspen center for physics
for hospitality during the course of this work.
O.A. and M.B. are also supported by the IRF Centers of
Excellence program, by the European RTN network HPRN-CT-2000-00122,
and by Minerva.
E.S. is supported in addition by the DOE (contract
DE-AC03-76SF00515 and OJI) and the Alfred P. Sloan Foundation.

\listrefs

\end